# Transit timings variations in the three-planet system: TOI-270

Laurel Kaye ,[1★] Shreyas Vissapragada ,[2] Maximilian N. Günther ,[3,4,5] Suzanne Aigrain,[1] Thomas Mikal-Evans ,[5] Eric L. N. Jensen ,[6] Hannu Parviainen ,[7,8] Francisco J. Pozuelos ,[9,10] Lyu Abe,[11] Jack S. Acton ,[12] Abdelkrim Agabi,[11] Douglas R. Alves ,[13] David R. Anderson ,[14,15] David J. Armstrong ,[14,15] Khalid Barkaoui ,[10,16] Oscar Barragán ,[1] Björn Benneke,[17] Patricia T. Boyd,[18] Rafael Brahm ,[19,20] Ivan Bruni ,[21] Edward M. Bryant ,[14,15] Matthew R. Burleigh ,[22] Sarah L. Casewell ,[22] David Ciardi,[23] Ryan Cloutier ,[24,25] Karen A. Collins ,[26] Kevin I. Collins ,[27] Dennis M. Conti ,[27] Ian J. M. Crossfield,[28] Nicolas Crouzet ,[29] Tansu Daylan ,[5,30] Diana Dragomir ,[31] Georgina Dransfield ,[32] Daniel Fabrycky ,[33] Michael Fausnaugh ,[5] Gábor Fűrész,[5] Tianjun Gan ,[34] Samuel Gill ,[15] Michaël Gillon ,[10] Michael R Goad ,[12] Varoujan Gorjian ,[35] Michael Greklek-McKeon ,[2] Natalia Guerrero,[3] Tristan Guillot ,[11] Emmanuël Jehin,[9] J. S. Jenkins ,[13,36] Monika Lendl ,[37] Jacob Kamler,[38] Stephen R. Kane ,[39] John F. Kielkopf ,[40] Michelle Kunimoto,[3,5] Wenceslas Marie-Sainte,[21] James McCormac ,[15] Djamel Mékarnia ,[11] Farisa Y. Morales ,[35] Maximiliano Moyano ,[41] Enric Palle,[7,8] Vivien Parmentier,[1] Howard M. Relles,[25] François-Xavier Schmider ,[11] Richard P. Schwarz ,[42] S. Seager ,[5,43,44] Alexis M. S. Smith ,[45] Thiam-Guan Tan ,[46] Jake Taylor ,[1] Amaury H. M. J. Triaud ,[32] Joseph D. Twicken ,[47,48] Stephane Udry ,[37] J. I. Vines ,[13] Gavin Wang ,[49,50] Peter J. Wheatley [14,15] and Joshua N. Winn [51]

*Affiliations are listed at the end of the paper*

## ABSTRACT

We present ground- and space-based photometric observations of TOI-270 (L231-32), a system of three transiting planets consisting of one super-Earth and two sub-Neptunes discovered by *TESS* around a bright (K-mag = 8.25) M3V dwarf. The planets orbit near low-order mean-motion resonances (5:3 and 2:1) and are thus expected to exhibit large transit timing variations (TTVs). Following an extensive observing campaign using eight different observatories between 2018 and 2020, we now report a clear detection of TTVs for planets c and d, with amplitudes of ∼10 min and a super-period of ∼3 yr, as well as significantly refined estimates of the radii and mean orbital periods of all three planets. Dynamical modelling of the TTVs alone puts strong constraints on the mass ratio of planets c and d and on their eccentricities. When incorporating recently published constraints from radial velocity observations, we obtain masses of $M_{\rm b} = 1.48 \pm 0.18\, M_{\oplus}$, $M_{\rm c} = 6.20 \pm 0.31\, M_{\oplus}$, and $M_{\rm d} = 4.20 \pm 0.16\, M_{\oplus}$ for planets b, c, and d, respectively. We also detect small but significant eccentricities for all three planets : $e_{\rm b} = 0.0167 \pm 0.0084$, $e_{\rm c} = 0.0044 \pm 0.0006$, and $e_{\rm d} = 0.0066 \pm 0.0020$. Our findings imply an Earth-like rocky composition for the inner planet, and Earth-like cores with an additional He/$H_2O$ atmosphere for the outer two. TOI-270 is now one of the best constrained systems of small transiting planets, and it remains an excellent target for atmospheric characterization.

**Key words:** planets and satellites: composition – planets and satellites: formation – planets and satellites: fundamental parameters.

## 1 INTRODUCTION

Over the course of its 2-yr mission from 2018 to 2020, the Transiting Exoplanet Survey Satellite (*TESS*) was tasked with detecting small planets orbiting nearby bright stars so that detailed follow-up studies of their characteristics and atmospheres can be performed (Ricker et al. 2014). Among the *TESS* discoveries, systems containing multiple transiting planets are especially interesting, because they enable comparative planetology. Of the ∼30 multiplanet systems *TESS* has discovered so far, there are several that contain planets near mean-motion resonances and thus are expected to display transit timing variations (TTVs). However, because *TESS* observes most target stars for relatively short time frames of about a month, these discoveries are relatively few, and to date, only three such systems have been reported (Kipping et al. 2019; Demory et al. 2020; Trifonov et al. 2021).

★ E-mail: laurel.kaye@physics.ox.ac.uk

TTVs give us an important way to probe the dynamics of a system and are especially powerful when combined with radial velocity (RV) observations. RV measurements rely on detecting the gravitational pull of a planet on its host star and form the bulk of higher-precision mass measurements for exoplanets. This makes them most sensitive to massive and closely orbiting planets. TTVs occur due to dynamical interactions between planets, which lead to deviations from their Keplerian orbits as predicted by Agol et al. (2005) and Holman & Murray (2005). They were first discovered for the system of Kepler-9 in 2010 and can be sensitive to detecting and characterizing small planets (Holman et al. 2010). Since this discovery, TTVs have been found quite commonly in Kepler data, and several tens of systems with significant TTVs are expected to be discovered in *TESS* light curves, such as the TOI-216 system (Dawson et al. 2019; Kane et al. 2019).

One such system with exciting potential for precise mass characterization through TTV and RV analysis is TOI-270, a nearby, bright, M3V-type star with three confirmed transiting exoplanets discovered by Günther, Pozuelos & Dittmann (2020, hereafter G20). The innermost planet, TOI-270 b, is a 1.25 $R_{\oplus}$ super-Earth with a period of 3.36 d. The outer two planets are sub-Neptunes, with radii of 2.42 $R_{\oplus}$ and 2.13 $R_{\oplus}$ and periods of 5.66 d and 11.38 d. The periods of planets d and c lie near the 2:1 mean-motion resonance, and those of planets c and b near the 5:3 resonance. Consequently, the planets are expected to interact dynamically with each other, resulting in significant TTVs, in particular, for planets c and d.

G20 discuss the amenability of this system for detailed characterization of the bulk density and atmospheric composition of its planets. This enables probing the formation history, molecular abundances, and potential signs of atmospheric or ocean loss with the James Webb Space Telescope (JWST) and the extremely large telescopes (ELTs). Constraints on the planet masses are vital to interpret the atmospheric spectra such observatories will gather in the future (see e.g. Batalha et al. 2019). However, the short (∼5 months) time span of the *TESS* and ground-based observations included in G20 was too small compared to the expected dominant time-scale of the TTVs (∼3 yr) to allow the TTVs to be detected.

We have now achieved a clear detection of TTVs by continuing to monitor the system from the ground and using *Spitzer*. *TESS* also re-observed the system in cycle 3, and the observations to date now span a large enough portion of the TTV 'super-period' to allow for a meaningful dynamical analysis of the system. However, TTV observations alone can be hard to interpret due to strong degeneracies between the orbital parameters. Relative uncertainties on pure-TTV mass determinations range from ∼5 to 100 per cent depending on the number of planets in the system, their orbital configuration, and the number, precision, and time span of the measured transit times (Agol & Fabrycky 2018).

In the meantime, Van Eylen et al. (2021, hereafter VE21) also monitored the TOI-270 system, using the Echelle Spectrograph for Rocky Exoplanet and Stable Spectroscopic Observations (ESPRESSO) to obtain RV measurements (Pepe et al. 2021). VE21 report masses of $1.58 \pm 0.26\,M_{\oplus}$, $6.14 \pm 0.38\,M_{\oplus}$, and $4.78 \pm 0.46\,M_{\oplus}$ for planets b, c, and d, respectively, classifying the inner planet as likely rocky, and the outer two as lower density sub-Neptunes. These results found no evidence of non-circular orbits and confirmed the status of planets c and d as outstanding targets for transmission spectroscopy and planet b as a good target for emission spectroscopy. The uncertainties on the masses reflect the small RV semi-amplitudes of the TOI-270 planets (2, 5, and 3 $\mathrm{m\,s^{-1}}$) and the limited number of RV measurements (∼80). This provides us with an opportunity to compare the results of an independent analysis of our TTV measurements with those reported by VE21 based on RV observations, use the RV results as priors in our analysis of the TTVs, and obtain refined estimates of the planets' masses and orbital parameters.

The combination of TTV and RV observations was first demonstrated for the Kepler-9 system by Holman et al. (2010) and has since been re-examined for the system multiple times (Borsato et al. 2014; Dreizler & Ofir 2014; Freudenthal et al. 2018; Borsato et al. 2019). Other systems that have been analysed jointly with RV and TTV data include K2-19 (Nespral et al. 2017), Kepler-19 (Malavolta et al. 2017), WASP-47 (Weiss et al. 2017), K2-24 (Petigura et al. 2018), and XO-6b (Garai et al. 2020). Some discrepancies have arisen between these two approaches (Weiss et al. 2013; Mills & Mazeh 2017); however, these seem to potentially be reconcilable with enough data points, as shown by Borsato et al. (2019). Nevertheless, these slight tensions highlight the importance of such joint studies for providing further elucidation on the biases and limits of these methods.

Precise planetary masses are essential for atmospheric characterization, particularly for resolving degeneracies in the interpretation of transmission spectra. Batalha et al. (2019) demonstrate that for small planets, robust atmospheric retrievals are possible only if the planet mass is known to a precision of 50 per cent or better, and that more precise mass estimates of 20 per cent are required to exploit the full sensitivity of JWST, given the latter's expected noise floor. The mass estimates presented by VE21 and reported in this work will thus have a direct impact on atmospheric studies of the system, starting with the ongoing *Hubble Space Telescope* (*HST*) observations (*HST*, program id GO-15814, PI Mikal-Evans). Our refined ephemeris and eccentricity constraints will also facilitate the scheduling of future transit and eclipse observations.

The rest of this paper is structured as follows. Section 2 details the photometric observations we obtained using eight distinct ground- and space-based telescope facilities, including data reduction and basic light curve extraction. Section 3 describes the light curve analysis, which we took care to perform in a consistent manner across all telescopes and gives the resulting estimates of the bulk parameters and individual transit times for each planet. Section 4 presents the dynamic analysis of TTVs using Differential Evolution Markov Chain Monte Carlo (DEMCMC). In Section 5, we discuss our improved mass estimates, system stability, potential atmospheric composition, and theories of planetary formation before concluding in Section 6.

## 2 OBSERVATIONS

This section details the observations and data reduction of each satellite and ground-based observatory. All observations are listed in Table 1.

Many of the ground-based observations discussed below were coordinated through the *TESS* Follow-up Observing Program (TFOP).[1] We used the TESS Transit Finder, which is a customized version of the Tapir software package (Jensen 2013), to schedule our transit observations. Unless otherwise noted, the photometric data were extracted using the AstroImageJ (`AIJ`) software package (Collins et al. 2017). Details of the facilities and instrumentation used are given in Table 2.

### 2.1 *TESS*

*TESS* first observed TOI-270 in short-cadence (2 min) observing mode between 2018 September 20 and 2020 December 17 in 27 d increments in sectors 3–5. *TESS* then recently re-observed the target in sectors 30 and 32 in camera 3, with 20-s cadence data[2] (see Fig. 1). Data were reduced and light curves extracted using the Science Processing Operations Center (SPOC) pipeline (Jenkins 2002; Jenkins et al. 2010; Smith et al. 2012; Stumpe et al. 2014; Jenkins et al. 2016; Jenkins 2017).

[1] https://tess.mit.edu/followup

**Table 1.** Observation log.

| TOI-270 | Dates | Telescope | Filter | Exposure time (seconds) | No. of exposures | Observation dur (minutes) | Transit coverage |
|---|---|---|---|---|---|---|---|
| **TOI-270 b-d** | | | | | | | |
| | 2018-2020 | TESS | TESS | 20-120 | 46874 | – | – |
| **TOI-270 b** | | | | | | | |
| | 2018-12-18 | PEST | Rc | 120 | 189 | 449 | Full |
| | 2019-01-14 | LCO | $i'$ | 90 | 172 | 160 | Full |
| | 2019-01-24 | LCO | – | 90 | 136 | 183 | Partial |
| | 2019-01-27 | LCO | – | 90 | 119 | 219 | Full |
| | 2019-09-03 | NGTS | NGTS | 10 | 8331 | 223 | Full |
| | 2019-09-30 | LCO | – | 90 | 115 | 164 | Full |
| | 2019-11-09 | LCO | – | 90 | 159 | 230 | Full |
| | 2019-11-16 | NGTS | – | 12 | 6811 | 208 | Full |
| | 2019-12-13 | LCO | – | 90 | 148 | 213 | Full |
| | 2020-09-13 | LCO | $i'$ | 90 | 128 | 198 | Full |
| | 2020-10-23 | LCO | $i'$ | 90 | 148 | 243 | Full |
| **TOI-270 c** | | | | | | | |
| | 2018-12-10 | LCO | $i'$ | 132 | 86 | 187 | Full |
| | 2019-01-07 | LCO | $i'$ | 51 | 207 | 215 | Full |
| | 2019-03-28 | ASTEP | Rc | 120 | 171 | 471 | Full |
| | 2019-04-19 | Spitzer | 4.5 | 2 | 10045 | 345 | Full |
| | 2019-07-30 | TRAPPIST-South | $z'$ | 10 | 511 | 170 | Full |
| | 2019-09-02 | TRAPPIST-South | $z'$ | 10 | 667 | 209 | Full |
| | 2019-09-08 | TRAPPIST-South | $z'$ | 10 | 742 | 232 | Full |
| | 2019-09-02 | LCO | – | 90 | 135 | 192 | Full |
| | 2019-09-13 | LCO | – | 90 | 94 | 134 | Partial |
| | 2019-09-30 | LCO | – | 90 | 119 | 170 | Partial |
| | 2019-10-17 | LCO | $i'$ | 90 | 132 | 189 | Full |
| | 2019-10-23 | TRAPPIST-South | $z'$ | 15 | 544 | 171 | Full |
| | 2019-11-03 | LCO | – | 90 | 159 | 230 | Full |
| | 2019-11-09 | TRAPPIST-South | $z'$ | 15 | 569 | 179 | Full |
| | 2019-11-09 | LCO | – | 90 | 160 | 229 | Full |
| | 2019-11-15 | LCO | – | 90 | 162 | 232 | Full |
| | 2019-11-26 | PEST | Rc | 130 | 91 | 210 | Partial |
| | 2019-12-02 | LCO | – | 90 | 162 | 232 | Full |
| | 2019-12-07 | Spitzer | 4.5 | 2 | 7643 | 263 | Full |
| | 2019-12-24 | Spitzer | 3.6 | 2 | 7510 | 262 | Full |
| | 2020-01-05 | Spitzer | 3.6 | 2 | 7380 | 258 | Full |
| | 2020-01-10 | Spitzer | 3.6 | 2 | 7462 | 260 | Full |
| | 2020-05-20 | ASTEP | Rc | 40 | 422 | 459 | Full |
| | 2020-05-25 | ASTEP | Rc | 40 | 345 | 373 | Full |
| | 2020-06-06 | ASTEP | Rc | 40 | 331 | 359 | Full |
| | 2020-10-25 | LCO | $i'$ | 90 | 211 | 328 | Full |
| | 2021-01-01 | LCO | $i'$ | 90 | | | Full |
| **TOI-270 d** | | | | | | | |
| | 2019-01-19 | LCO | $i'$ | 50 | 182 | 156 | Partial |
| | 2019-02-23 | LCO | $g'$ | 96 | 123 | 192 | Partial |
| | 2019-06-16 | Spitzer | 4.5 | 2 | 10851 | 361 | Full |
| | 2019-08-24 | TRAPPIST-South | $z'$ | 10 | 369 | 115 | Partial |
| | 2019-09-27 | LCO | – | 90 | 161 | 231 | Full |
| | 2019-09-27 | NGTS | NGTS | 10 | 5037 | 273 | Full |
| | 2019-10-08 | LCO | – | 90 | 132 | 192 | Full |
| | 2019-10-20 | LCO | – | 90 | 115 | 165 | Partial |
| | 2019-10-31 | LCO | – | 90 | 97 | 138 | Partial |
| | 2019-11-23 | TRAPPIST-South | $z'$ | 10 | 630 | 197 | Full |
| | 2019-11-23 | NGTS | NGTS | 10 | 9167 | 328 | Full |
| | 2019-11-23 | CHAT | – | – | – | – | Partial |
| | 2019-12-04 | Spitzer | 4.5 | 2 | 7600 | 253 | Full |
| | 2020-01-07 | Spitzer | 4.5 | 2 | 7583 | 261 | Full |
| | 2020-07-19 | LCO | $i'$ | 90 | 69 | 106 | Partial |
| | 2020-09-02 | LCO | $i'$ | 90 | 124 | 190 | Partial |

**Table 2.** Facilities used for ground-based follow-up observations.

| Observatory | Location | Aperture (m) | Pixel scale (arcsec) | FOV (arcmin) |
|---|---|---|---|---|
| Antarctic Search for Transiting ExoPlanets (ASTEP) | Concordia Station, Antarctica | 0.4 | 0.93 | 63 × 63 |
| Chilean-Hungarian Automated Telescope (CHAT) | Las Campanas Observatory, Chile | 0.7 | 0.6 | 21 × 21 |
| Las Cumbres Observatory Global Telescope (LCOGT) | Chile, South Africa, Australia | 1.0 | 0.39 | 26 × 26 |
| Next-Generation Transit Survey (NGTS) | Paranal, Chile | 0.2 | 4.97 | 170 × 170 |
| Perth Exoplanet Survey Telescope (PEST) | Perth, Australia | 0.3 | 1.2 | 31 × 21 |
| TRAPPIST-South | La Silla, Chile | 0.6 | 0.64 | 22 × 22 |

### 2.2 *LCO*

TOI-270 was observed by the Las Cumbres Observatory Global Telescope (LCOGT) (Brown et al. 2013) 1-m telescope network as part of ground-based follow-up as reported in G20, through subsequent observations as part of the *TESS* Follow-up Observing Program (TFOP), and through an additional program with PI Parviainen over semesters 2019B, 2020B, and 2021A. The 4096 × 4096 LCOGT SINISTRO cameras have an image scale of 0.″389 per pixel, resulting in a 26′ × 26′ field of view. The images were calibrated by the standard LCOGT `BANZAI` pipeline (McCully et al. 2018).

The LCOGT observations led by Parviainen cover six full and one partial TOI-270c transits and one full and two partial TOI-270d transits. The observations were carried out with SINISTRO cameras using SDSS-ip filter with 60-s exposure times and slight defocussing to make sure neither the target star nor any of the comparison stars saturate. As with the TFOP LCOGT observations, the images were calibrated by the BANZAI pipeline. The photometry was done using a custom photometry pipeline developed for the MuSCAT2 instrument (Narita et al. 2019; Parviainen et al. 2019), and an initial transit modelling was carried out using the `PyTransit` package (Parviainen 2015).

### 2.3 *TRAPPIST*-South

TRAPPIST-South at ESO-La Silla Observatory in Chile is a 60-cm Ritchey–Chretien telescope, which has a thermoelectrically cooled $2K \times 2K$ FLI Proline CCD camera with a field of view of 22′ × 22′ and pixel scale of 0.65 (Jehin et al. 2011; Gillon et al. 2013). We carried out a set of five full-transit observations of TOI-270 c and one full and one partial transits for TOI-270 d.

### 2.4 *ASTEP*

ASTEP is 40-cm telescope installed at the Concordia Station, Dome C, Antarctica (Guillot et al. 2015). ASTEP is a Newtonian telescope equipped with a five-lens Wynne coma corrector and a $4k \times 4k$ front-illuminated FLI Proline KAF 16810E CCD with a 16-bit dynamic. The camera has an image scale of 0.92” pixel$^{-1}$ resulting in a 63′ × 63′ corrected field of view (Abe et al. 2013).

Due to the extremely low data transmission rate at the Concordia Station, the data are processed on-site using an automated IDL-based pipeline producing data files that contain the star’s flux computed through various fixed circular apertures radii, so that optimal light curves can be extracted (Mékarnia et al. 2016). TOI-270 was observed several times between 2019 March and April and between 2020 April and July . For this study, we selected a set of four full-transit observations of TOI-270 c conducted under good weather conditions with temperatures ranging from −70°C to −60°C.

### 2.5 *NGTS*

The Next-Generation Transit Survey (NGTS; Wheatley et al. 2018) consists of 12 independently operated robotic telescopes, each with a 20-cm diameter and an 8 square-degree field of view, and is located at the ESO Paranal Observatory in Chile. The NGTS telescopes achieve sub-pixel-level guiding through the use of the DONUTS auto-guiding algorithm (McCormac et al. 2013). Observing a single bright star simultaneously with multiple NGTS telescopes has been shown to significantly improve the photometric precision achieved (Bryant et al. 2020; Smith et al. 2020)

Two transits of TOI-270b were observed on 2019 September 3 and 2019 November 16 using eight and seven telescopes, respectively. A total of 15 142 images were obtained across the two nights. A further two transits of TOI-270d were observed on 2019 September 27 and 2019 November 23, using four and six telescopes. A total of 14 204 images were taken across these two nights. All the NGTS observations of TOI-270 were performed using the custom NGTS filter (520–890 nm) and with an exposure time of 10 s.

A custom aperture photometry pipeline (Bryant et al. 2020) was used to reduce the images from all four nights of NGTS observations. This pipeline uses the `SEP` library (Bertin & Arnouts 1996; Barbary 2016) for source extraction and photometry. During the reduction comparison stars that are similar to TOI-270 in terms of brightness, colour, and CCD position are automatically identified using *Gaia* DR2 (Gaia Collaboration 2016, 2018).

### 2.6 *PEST*

A single transit of planet c was observed by the Perth Exoplanet Survey Telescope (PEST). A custom pipeline based on `C-Munipack`[3] was used to calibrate the images and extract the differential photometry.

[2]Thanks to the *TESS* GI programs G03196 (PI Günther), G03278 (PI Mayo), G03106 (PI Kane), and G03274 (PI Cloutier)

[3]http://c-munipack.sourceforge.net

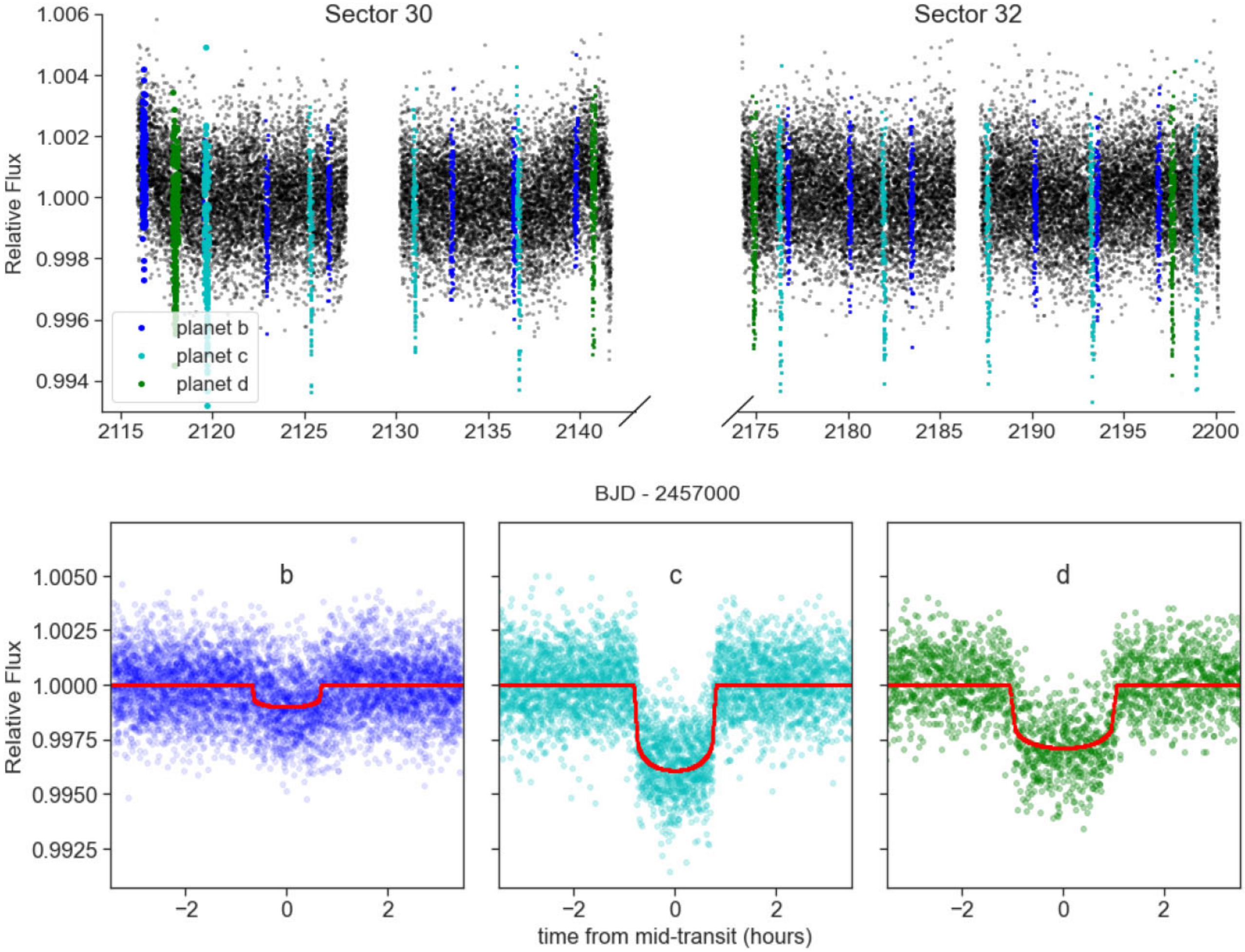

**Figure 1.** The top plot shows *TESS* sectors 30 and 32 in PDC-SAP flux, with transits highlighted for each planet. The bottom panels show phase-folded transits for the respective planets, with the red showing the best-fitting model.

## 2.7 *Spitzer*

Soon after the TOI-270 planet candidates were announced by the *TESS* project, we identified the system as an intriguing target for transit observations with the *Spitzer Space Telescope*. We observed five transits of TOI-270 c and three transits of TOI-270 d with *Spitzer*, four transits in each of two programs (GO-14084 and 14325, Crossfield et al. 2018; Parmentier et al. 2019). Our *Spitzer* observations provided coverage of these planets' transits in each of the 3.6-µm and 4.5-µm channels of the Infrared Array Camera (IRAC) instrument (Fazio et al. 2004). We planned our observations following standard practices for precise *Spitzer* transit photometry. In particular, an initial peak-up observation placed the target on the IRAC detector's 'sweet spot' to minimize instrumental variations.

Because of the bright target star both channels of *Spitzer* photometry used subarray-mode observations, which consist of multiple sets of 64 quick subarray frames. All our *Spitzer* observations used 2-s subarray integrations, and raw and calibrated data products (processed with the standard *Spitzer* data calibration pipeline) are publicly available through the Spitzer Heritage Archive.[4]

[4] https://sha.ipac.caltech.edu/applications/Spitzer/SHA/

## 3 LIGHT-CURVE ANALYSIS

When analysing the light curves from the different telescopes, our aim was to get as close as possible to a fully self-consistent analysis of all the available photometric data for each planet. However, this quickly results in an unfeasibly large number of data points and free parameters, as each transit has a specific mid-transit time, and each telescope a specific set of limb-darkening and noise model parameters, in addition to the bulk transit parameters that are common to all data sets. We therefore reached a compromise, whereby the light-curve analysis was done sequentially, as follows.

We first fit the bulk parameters using the *TESS* cycle 1 data, ignoring the TTVs (which are negligible on this time-scale). In the case of planets c and d, we then used the resulting linear ephemeris (period and time of conjugation) and bulk parameters as inputs to analyse the *Spitzer* data. This results in significantly tighter constraints on the bulk planet parameters. We then proceed to fit each transit from every instrument in turn, fixing the bulk parameters[5] at the *Spitzer*-derived values, and varying only the

[5] Early in the study, we tried allowing the bulk transit parameters to vary when fitting for the TTVs, using the *Spitzer*-derived constraints as priors. This did not appreciably increase the uncertainties on the transit times and resulted in

**Table 3.** Planetary parameters.

| Parameter | Source | | |
|---|---|---|---|
| **Planet b** | **G20** | **Spitzer** | **TESS** |
| $T_0$ | $1461.01464^{+0.00084}_{-0.00093}$ | | $1461.01556^{+0.0005859}_{-0.0005894}$ |
| Period (days) | $3.36008 \pm 0.00007$ | | $3.36016 \pm 0.000004$ |
| $R_p/R_*$ | $0.0300^{+0.0015}_{-0.0011}$ | | $0.0307^{+0.0012}_{-0.00091}$ |
| $\cos i$ | $0.024^{+0.024}_{-0.015}$ | | $0.0267^{+0.0256}_{-0.0184}$ |
| $a/R_*$ | $17.48251^{+3.97329}_{-1.37537}$ | | $17.10761^{+3.52558}_{-1.85761}$ |
| **Planet c** | | | |
| $T_0$ | $1463.08481 \pm 0.00025$ | $1463.08056^{+0.00039}_{-0.00040}$ | |
| Period (days) | $5.66017 \pm 0.00004$ | $5.66057 \pm 0.00001$ | $5.66057 \pm 0.00001$ |
| $R_p/R_*$ | $0.05825^{+0.00079}_{-0.00058}$ | $0.05595 \pm 0.00171$ | $0.05781^{+0.00067}_{-0.00058}$ |
| $\cos i$ | $0.0083^{+0.0073}_{-0.0051}$ | $0.01393^{+0.00230}_{-0.00282}$ | $0.00698^{+0.00657}_{-0.00476}$ |
| $a/R_*$ | $27.005^{+1.677}_{-0.591}$ | $25.5690 \pm 0.3062$ | $27.443^{+0.512}_{-1.253}$ |
| **Planet d** | | | |
| $T_0$ | $1469.33834^{+0.00052}_{-0.00046}$ | $1469.33823 \pm 0.00032$ | |
| Period (days) | $11.38014^{+0.00011}_{-0.00010}$ | $11.37948 \pm 0.00002$ | $11.37958 \pm 0.00003$ |
| $R_p/R_*$ | $0.05143 \pm 0.00074$ | $0.04914 \pm 0.00158$ | $0.05181^{+0.00048}_{-0.00078}$ |
| $\cos i$ | $0.0054^{+0.0021}_{-0.0027}$ | $0.0062^{+0.0019}_{-0.0026}$ | $0.0062^{+0.0028}_{-0.0043}$ |
| $a/R_*$ | $41.5627^{+0.864}_{-0.691}$ | $41.7437 \pm 0.5274$ | $41.776^{+1.353}_{-0.359}$ |

*Note.* Stellar parameters and planetary priors are drawn from G20 and are shown in the first column. $T_0$ is reported in days ($BJD_{TDB}$) − 2 457 000. The middle column shows planetary parameters from fitting *Spitzer* transits as described above. The third column shows *TESS* fits incorporating all available data to date from sectors 3–5, 30, and 32. Limb-darkening coefficients are not shown, as they were fit using LDTk (Parviainen & Aigrain 2015) for each individual observatory and filter used. Timing fits were performed by fixing planetary parameters to the *Spitzer*-derived values for planet c, and for *TESS* all-sector values for planets b and d.

parameters of the noise model, the mid-transit time, and the limb-darkening parameters.

In the case of planet b, in the absence of *Spitzer* observations, we proceed in the same manner but relied on the *TESS* cycle 1 and cycle 3 data to derive the bulk parameters and mean period used as priors in the individual fits.

This sequential, Bayesian approach enables us to keep the computational cost of each fit manageable, while ensuring the overall self-consistency of the results. The final planetary parameters are reported in Table 3, and the mid-transit times for each transit (excluding those that were discarded because the fits were deemed unreliable, as discussed below) in Tables A1–A3.

### 3.1 Model implementation

The light-curve modelling was implemented in PYTHON. All transit modelling was done using the BATMAN package (Kreidberg 2015). The parameters of the transit model implemented in BATMAN are the orbital period $P$, the time of transit centre $T_0$, the planet-to-star radius ratio $R_p/R_\star$, the system scale $a/R_\star$ (where $a$ is the orbital semimajor axis), the orbital inclination $i$, the orbital eccentricity $e$ and the longitude of periastron $\omega$, and the quadratic limb-darkening coefficients $u_1$ and $u_2$. In the present work, we fixed the eccentricity to zero, an assumption that is reasonable, given the small eccentricity measured in RV by VE21 and is ultimately validated by our dynamical analysis of the TTVs (see Section 4). When fitting, we chose to fit for $R_p$, $R_\star$, and $M_\star$, which are then used to obtain $R_p/R_\star$ and $a/R_\star$ (through Kepler's third law). We used the Limb Darkening Toolkit (LDTK) of Parviainen & Aigrain (2015) to compute priors over the limb-darkening coefficients $u_1$ and $u_2$, using the stellar parameters from G20 and the relevant instrument/filter throughput as inputs. We used uninformative, wide uniform priors over all the other parameters (except for inclination where we used a uniform prior in $\cos i$), except where explicitly stated otherwise below.

To account for correlated noise in the *TESS* and ground-based light curves, we used a Gaussian Process (GP) with a squared exponential covariance function:

$$k(t, t') = A \exp\left[-\frac{(t-t')^2}{2\tau}\right], \tag{1}$$

where $A$ is the variance and $\tau$ the (squared) characteristic time-scale, or metric, of the GP. All GP calculations in this work were done using the GEORGE package (Ambikasaran et al. 2014). We used log uniform priors over the logarithm of the GP hyper-parameters $A$ and $\tau$.

When performing optimization, we first used the MINIMIZE function of the SCIPY.OPTIMIZE module to determine initial input parameters and explored the posterior distribution over the parameters of our models using Markov Chain Monte Carlo (MCMC) implemented with the EMCEE package (Foreman-Mackey et al. 2013, 2019). We evolved chains of 40 walkers with a thinning factor of 15 with a conservative removal of the first 25 per cent of steps as burn-in. Convergence was achieved when the estimated autocorrelation time $\tau_{autocor}$ (checked every 200 steps) was changed by less than 1 per cent and the chain was longer than 200 times the autocorrelation time, resulting in chain lengths of about 100 000. Parameter values were taken to be the median of the posterior distribution, the with errors much longer convergence times, given the larger number of parameters to fit, so we opted to fix the bulk parameters instead when fitting for the TTVs.

reported as the 68 per cent confidence interval from the 16th to the 84th percentile.

### 3.2 *TESS*

Our starting point for the *TESS* data is the 2-min cadence Pre-search Data Conditioning- Simple Aperture Photometry (PDC-SAP) light curves, in which most systematic effects have already been corrected (Smith et al. 2012; Stumpe et al. 2012; Stumpe et al. 2014). None the less, we checked for any residual correlated noise using the out-of-transit data in two distinct ways. First, we computed the auto-correlation function (ACF) of the out-of-transit light curve and confirmed that it behaves as expected for white Gaussian noise, namely the ACF is within a standard deviation of zero for all non-zero lags. Secondly, we tried modelling the out-of-transit light curve using a GP, as described in Section 3.1, and found that this resulted in a very small best-fitting value for $A$. In the remainder of our analysis of the *TESS* data, we do include a GP term, primarily for consistency with our later analysis of the ground-based data, but we fix the hyper-parameters at the best-fitting values obtained on the out-of-transit cycle 1 (sectors 3–5) data. We note that essentially identical results would be obtained by either varying the hyper-parameters while fitting the transits or not including a GP at all for the *TESS* data.

We then proceed to model the transits, initially in the cycle 1 data only, as the effects of the TTVs within the limited time-frame of those data are negligible, enabling a global analysis of the transits of each planet in turn, assuming a linear ephemeris. We start by extracting a segment of the light curve lasting three transit durations around the predicted centre of each transit, using the linear ephemeris from G20. We then fit the data in and around all the transits for each planet, keeping the GP hyper-parameters fixed, and varying the transit parameters only. The results of this analysis are shown in Table 3 and are in good agreement with the discovery paper, as expected.

Finally, to measure TTVs, we fix the planetary parameters to the values derived from the *Spitzer* data (for planets c and d, see Section 3.3 for details) and from *TESS* cycle 1 and cycle 3 (for planet b) and fit each transit individually, varying only the time of the transit centre.

### 3.3 Spitzer

With its IRAC in this warm viewing mode, which under-samples the point spread function (PSF), *Spitzer* spacecraft wobble and jitter causes the PSF to move across pixels resulting in significant correlated noise (Ingalls et al. 2012). This is a well-characterized problem, and several methods have been used previously in order to remove these effects (Lanotte et al. 2014; Shporer et al. 2014). In this work, in order to reduce the level of systematic error seen in warm *Spitzer* data, we followed the process of pixel-level decorrelation (PLD) as described by Deming et al. (2015) with the new addition of principal component analysis (PCA). In summary, PLD employs a 3 x 3 grid of pixels that captures the stellar image over a time series. Each pixel produces its own time series across the duration of the observation, which accounts for any movement of the PSF of the star during that time (usually less than 1 pixel drift). The total light curve can then be expressed as a linear combination of a transit eclipse, the normalized pixel time series, and linear or higher-order trends in time. If the light curve we are interested in is a function of these pixel time series, then PCA enables us to reduce the dimensionality of a data set by looking for uncorrelated components that contribute to the variance seen. In the end, this results in a function for the total light over time in the form:

$$\Delta S^t = \sum_{i=1}^{3} c_i \hat{C}_i^t + D_{\rm transit}(t) + b_0 + b_1 t, \qquad (2)$$

where $\Delta S^t$ is the brightness of the star over time, $c_i \hat{C}_i^t$ are the principal component time series regularized at each time point (and thus do not contain the transit information), $D_{\rm transit}$ is the depth of the transit at any given time point (holding a value of 1 out of transit), and $b_0$ and $b_1$ are terms describing the constant and linear offsets. Higher-order polynomial offsets were not found to impact the transit fit in the case of Deming et al. (2015), and we found that to be the case in our work as well.

We therefore proceeded as follows: For each transit, we first perform a normalization using out-of-transit flux. Then, instead of using all 9 pixel time series, which are likely degenerate, we performed PCA and found that using three components explained >95 per cent of the variance. Reducing this to two components resulted in a slight worsening of the eventual transit fit, and adding a fourth component did not produce a measurable change. We then constructed a model using the equation above with the three principal components of the pixel series, a BATMAN light curve as described in Section 3.1, and a linear and constant offset. This resulted in a fit that is exemplified in Fig. 2 and Fig. A1. We then optimized for $c_i\, b_0\, b_1$ and all planetary parameters described above using wide, uniform priors. This conferred a slight improvement on the bulk planetary parameters derived from *TESS* as shown in Table 3. We adopted these as the fixed planetary parameters for the remainder of the ground-based fitting described below for planets c and d (no *Spitzer* observations were taken of planet b). We further returned to *TESS* data for planets c and d and fixed the bulk planetary parameters to *Spitzer*-derived values, while fitting for transit time, GP hyper parameters, and limb darkening.

### 3.4 Ground-based observatories

LCO observations were next fit in an identical manner to *TESS* using GP normalization simultaneously with an MCMC. First, to test whether ground-based observations could further constrain the bulk planetary parameters achieved by *Spitzer*, each transit was individually fit for all planetary parameters including mid-transit time using posterior distributions from the *Spitzer* fitting as priors. However, such fits for planetary parameters proved to be significantly less constraining than the *Spitzer* fit; therefore, in a second step, the bulk planetary parameters were held fixed at *Spitzer* fit values (or *TESS* fit parameters for planet b), and fitting was performed only for mid-transit time and GP hyperparameters, using wide uninformative uniform priors, and for limb-darkening parameters using LDTk. At first, this resulted in GP over-fitting, with a very small GP time-scale tau, as well as likelihood splitting around two values of tau (a high and a low, with the low being associated with over-fitting), and thus two local likelihood maxima. This was resolved by running a 'burn-in' phase of 600 steps and then selecting the highest likelihood walker value to initialize the subsequent run to convergence, which invariably resulted in the higher value of tau being selected.

Other ground-based observatory data including NGTS, TRAPPIST, PEST, and CHAT were carried out in an identical manner, with limb-darkening coefficients fit using LDTk with each filter's respective transmission functions used as input. In all cases, for ground-based observatories, *Spitzer* data fitting resulted in more constrained planetary parameters, so these were again held fixed,

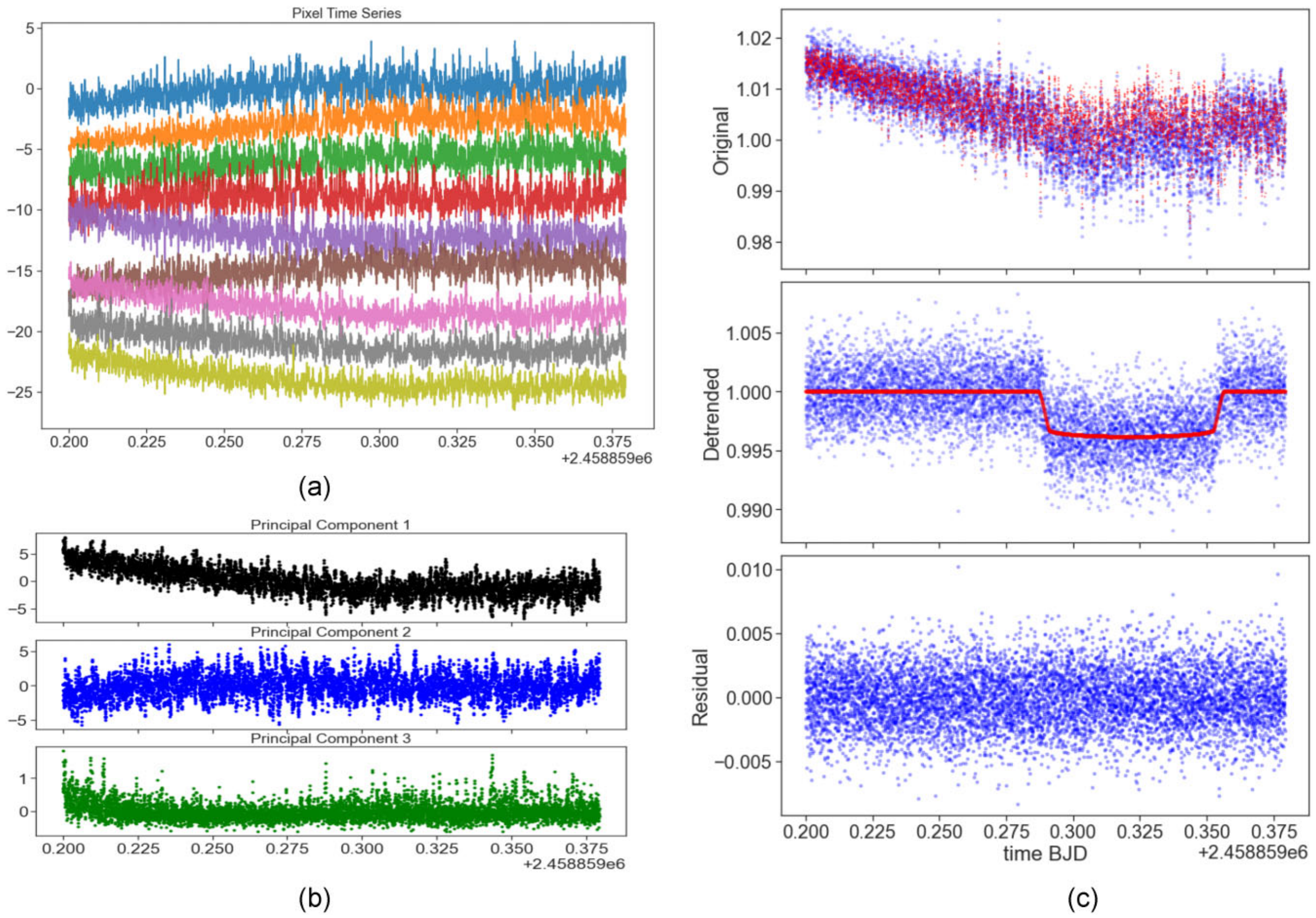

**Figure 2.** Pixel-level decorrelation (PLD), which is used to remove correlated noise from *Spitzer* light curves (Deming et al. 2015), is shown here for a transit of planet c. In (a), the regularized pixel time series are shown, which come from a 3 × 3 grid of pixels. Because of the regularization at each time point, these time series do not show a transit. From these, we extract the three principal components shown in (b), the first of which follows the most obvious trend of the original light curve, which can be seen in the top panel of (c) in blue. Also shown in the top panel of (c) in red is the best-fitting noise model and solution to equation (2). In the middle panel of (c), the noise is removed and the best-fitting transit model is shown in red. The bottom panel shows the final residuals after fitting.

and only the timing offsets, limb darkening, and GP hyperparameters were fit.

If, upon visual inspection, the fit resulted in an obviously incorrect transit centre or clear residuals showing correlated noise, the transit was discarded. In all cases, this was due to either too few data points or to significant residual correlated noise (often attributable to airmass) as can be seen in Figs A2 and A3. A clear anticorrelated trend of TTVs is shown between the two outer planets in Fig. 3 with an amplitude of around 10 min, which is on the lower end of the range of the predictions made by G20. We perform dynamical analysis on these timing offsets in further detail in the following section.

## 4 DYNAMICAL ANALYSIS

### 4.1 TTV retrieval

Our long-term follow-up campaign revealed a near-linear ephemeris for TOI-270 b and substantial, anticorrelated deviations from linear ephemerides for TOI-270 c and d. We modelled the transit timing data for all three planets using the `ttvfast` code (Deck et al. 2014). Given the planetary and stellar masses along with the osculating planetary orbital elements (using the orbital periods instead of the semimajor axes) at the beginning of the integration, `ttvfast` rapidly forward-models the transit timings, which can then be compared to observations. Rather than fitting directly in these parameters, we chose a basis that allows us to speed up computations and avoid bias. First, rather than fitting for the eccentricity $e$ and argument of periastron $\omega$ separately, we fit in the $(\sqrt{e}\cos\omega, \sqrt{e}\sin\omega)$ basis. This avoids the pitfalls of fitting for periodic parameters while naturally admitting a uniform prior on $e$ and $\omega$ due to the constant Jacobian (Eastman, Gaudi & Agol 2013). Secondly, we used the time of first transit as a convenient reparametrization of the mean anomaly (e.g. Jontof-Hutter et al. 2016; Libby-Roberts et al. 2020). Thirdly, for the purposes of the dynamical modelling, we fixed the inclinations and longitudes of ascending node to 90° and 0°, respectively. This is commonly done to reduce dimensionality in TTV modelling, as the signal has only a higher-order dependence on the mutual inclinations (e.g. Lithwick, Xie & Wu 2012; Nesvorný & Vokrouhlický 2014), which are known to be small for multiplanet systems in general (Fabrycky et al. 2014; Gilbert & Fabrycky 2020) and for the multitransiting TOI-270 specifically (G20). All in all, we fit five parameters ($M_p/M_\star$, $P$, $\sqrt{e}\cos\omega$, $\sqrt{e}\sin\omega$, and $T_0$) for each of the three planets, giving a total of 15 fit parameters. Orbital elements were osculating and defined at the simulation start time BJD $= 2458381$ using Jacobi coordinates.

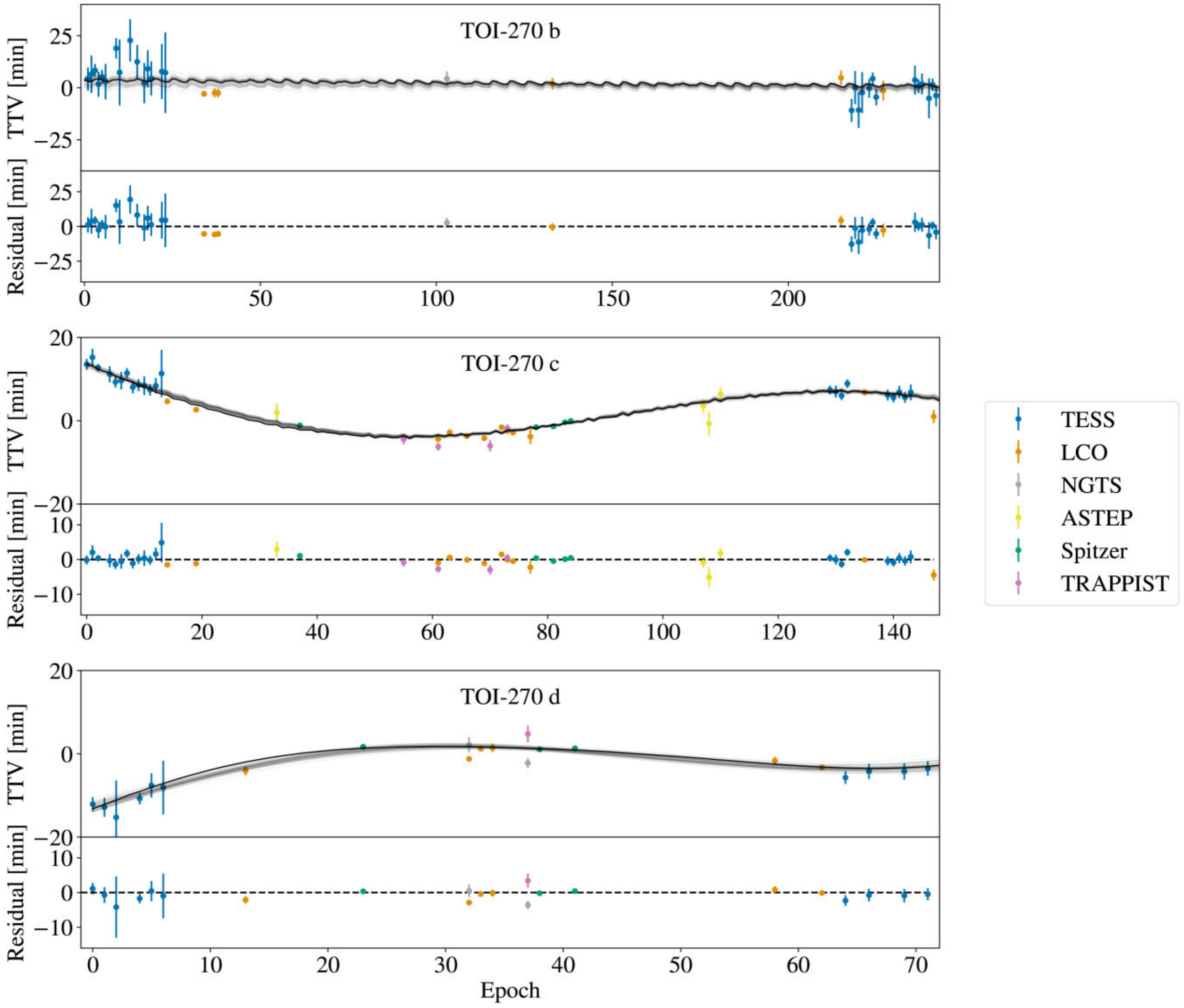

**Figure 3.** Results of the TTV analysis. In the top panels, the timing data are shown with a linear ephemeris (determined via least-squares) removed. We show the maximum a posteriori probability (MAP) model with dark blue curves, and 100 random samples from the posterior are plotted in light blue to give a sense for the uncertainty. In the bottom panels, the residuals from the MAP model are given, with the black dashed line indicating a perfect fit.

With our forward model defined, we proceeded to find the model parameters with DEMCMC, implemented in the v3.0 of the EMCEE package (Foreman-Mackey et al. 2013, 2019). We used Gaussian priors on the planet mass ratios from VE21, wide uniform priors on all other parameters, and these priors are noted in Table 4, labelled as 'TTV + RV'. We also ran a dynamical retrieval with wide uniform priors on the planet mass ratios to assess the degree to which the TTVs alone could constrain the system dynamics; those priors are noted in Table 4 as well and labelled 'TTV-only'. In this fit, we used only linear ephemerides for planet b, fixing the eccentricity to 0 and the mass to the value from VE21, to isolate the dominant impact of the 2:1 resonance for planets c and d and improve convergence. When computing the likelihood, we assumed that the data were drawn from a Student $t$ distribution with 2 degrees of freedom rather than a Gaussian. Previous studies of *Kepler* TTV data found that this distribution tends to better fit the timing residuals than a Gaussian (Jontof-Hutter et al. 2016; MacDonald et al. 2016), and the Student $t$ distribution can also help retrievals on large, heterogeneous data sets remain resilient to outliers (Agol et al. 2020). While Agol et al. (2020) use a Student $t$ distribution with 4 degrees of freedom for analysing TTVs in the TRAPPIST-1 system, we choose to adopt 2 degrees of freedom following Jontof-Hutter et al. (2016), whose Kepler residual histogram with more data points displays a higher signal to noise. We verified that these choices did not skew our results by first fitting the TTVs of the well-characterized three-planet system Kepler-51; we used the same fitting basis as Libby-Roberts et al. (2020) and reproduced their posteriors to well within $1\sigma$ on each of the model parameters.

We first optimized the model using the Powell minimizer implemented in scipy.optimize.minimize, and then initialized 48 walkers for our DEMCMC in a Gaussian ball around this starting point. We ran each chain for $5 \times 10^6$ steps to burn-in the sampler and then proceeded to run for $2 \times 10^7$ steps. This rather long chain length

**Table 4.** Orbital parameters obtained from TTV analysis.

| Parameter | Prior (TTV + RV) | Posterior 1 (TTV + RV; adopted) | Prior 2 (TTV only) | Posterior 2 (TTV only) | Units |
|---|---|---|---|---|---|
| $M_b$ | $\mathcal{N}(1.58, 0.26)$ | $1.48^{+0.19}_{-0.18}$ | 1.58 (fixed) | – | $M_\oplus$ |
| $P_b$ | $\mathcal{U}(3.35, 3.37)$ | $3.35992^{+0.00005}_{-0.00004}$ | $\mathcal{U}(3.35, 3.37)$ | $3.35999^{+0.00005}_{-0.00007}$ | Days |
| $\sqrt{e_b}\cos\omega_b$ | $\mathcal{U}(-0.7, 0.7)$ | $0.007^{+0.085}_{-0.067}$ | 0 (fixed) | – | – |
| $\sqrt{e_b}\sin\omega_b$ | $\mathcal{U}(-0.7, 0.7)$ | $-0.005^{+0.061}_{-0.059}$ | 0 (fixed) | – | – |
| $T_{0,\,b}$ | $\mathcal{U}(2.65, 2.81)$ | $2.72985^{+0.00111}_{-0.00129}$ | $\mathcal{U}(2.65, 2.81)$ | $2.7301^{+0.0012}_{-0.0013}$ | Days |
| $M_c$ | $\mathcal{N}(6.14, 0.38)$ | $6.20^{+0.31}_{-0.30}$ | $\mathcal{U}(0, 50)$ | $4.5^{+1.7}_{-1.2}$ | $M_\oplus$ |
| $P_c$ | $\mathcal{U}(5.65, 5.67)$ | $5.66051 \pm 0.00004$ | $\mathcal{U}(5.65, 5.67)$ | $5.66052 \pm 0.00005$ | Days |
| $\sqrt{e_c}\cos\omega_c$ | $\mathcal{U}(-0.7, 0.7)$ | $0.008^{+0.019}_{-0.017}$ | $\mathcal{U}(-0.7, 0.7)$ | $-0.025^{+0.014}_{-0.015}$ | – |
| $\sqrt{e_c}\sin\omega_c$ | $\mathcal{U}(-0.7, 0.7)$ | $0.077^{+0.006}_{-0.007}$ | $\mathcal{U}(-0.7, 0.7)$ | $0.0700^{+0.0058}_{-0.0055}$ | – |
| $T_{0,\,c}$ | $\mathcal{U}(2.82, 2.86)$ | $2.8417 \pm 0.0003$ | $\mathcal{U}(2.82, 2.86)$ | $2.84157^{+0.00029}_{-0.00029}$ | Days |
| $M_d$ | $\mathcal{N}(4.78, 0.46)$ | $4.20 \pm 0.16$ | $\mathcal{U}(0, 50)$ | $3.62^{+0.47}_{-0.47}$ | $M_\oplus$ |
| $P_d$ | $\mathcal{U}(11.36, 11.4)$ | $11.38194 \pm 0.00010$ | $\mathcal{U}(11.36, 11.4)$ | $11.38165^{+0.00045}_{-0.00033}$ | Days |
| $\sqrt{e_d}\cos\omega_d$ | $\mathcal{U}(-0.7, 0.7)$ | $-0.021^{+0.071}_{-0.041}$ | $\mathcal{U}(-0.7, 0.7)$ | $-0.090^{+0.021}_{-0.017}$ | – |
| $\sqrt{e_d}\sin\omega_d$ | $\mathcal{U}(-0.7, 0.7)$ | $0.047^{+0.024}_{-0.036}$ | $\mathcal{U}(-0.7, 0.7)$ | $0.016^{+0.026}_{-0.021}$ | – |
| $T_{0,\,d}$ | $\mathcal{U}(8.66, 8.70)$ | $8.68021 \pm 0.00052$ | $\mathcal{U}(8.66, 8.70)$ | $8.67964^{+0.00071}_{-0.00073}$ | Days |

*Note.* Orbital elements are osculating and defined using Jacobi coordinates at the simulation start time BJD $= 2458381$.

was required to capture more than 50 autocorrelation lengths in the ($\sqrt{e}\cos\omega$, $\sqrt{e}\sin\omega$) parameters, whereas the other parameters converged comparatively rapidly. The posterior distributions for our model parameters are summarized in Table 4 and visualized for the TTV + RV fit in Fig. A4. For ease of comparison to VE21, we transformed the planet-to-star mass ratios back to the planetary masses in this figure and table.

The results of our TTV-only retrieval agree reasonably well with the analysis of VE21. The mass of planet c is concordant to within $1\sigma$ considering the uncertainty on both measurements, but there is a slight tension at the $1.5\sigma$ level between the results for planet d, with TTVs suggesting a slightly lower mass. However, the TTV-only masses are less precise than the RV masses. Still, the reasonable agreement between independently derived masses suggests that both experiments have good control over systematic effects, and that the masses are realistic. Additionally, from our TTV-only posteriors, we obtain a mass ratio of $M_d/M_c = 0.80^{+0.17}_{-0.14}$, in excellent agreement with VE21, who measured $M_d/M_c = 0.78 \pm 0.09$.

To get the full benefit of both data sets, we adopt our TTV + RV dynamical solution in Table 4, which is shown with the TTV data in Fig. 3. The precision of the mass constraints is modestly increased for planets b and c and is greatly improved for planet d. This is because the TTVs place strong constraints on the covariance between $M_c$ and $M_d$, so prior information on one or both of these masses results in a very precise posterior. We are further able to constrain the eccentricities and arguments of periastron for all three planets. The TTV data suggest that the eccentricities of the TOI-270 planets are quite low, with $e_b < 0.024$ (95th percentile), $e_c = 0.00619^{+0.00092}_{-0.00088}$, and $e_d < 0.011$ (95th percentile). While we can only place an upper limit on the inner and outer planets' eccentricities, the eccentricity of TOI-270 c is detected. Moreover, the eccentricity constraints are not too affected by the choice of mass prior: for the uniform prior retrieval, we obtained $e_c = 0.00575^{+0.00097}_{-0.00093}$, and $e_d < 0.015$ (95th percentile). More precise constraints on the masses and eccentricities will await further sampling of the TTV curve by *TESS* and other instruments.

### 4.2 Long-term stability

Using these dynamical constraints, we performed a suite of tests aimed at characterizing the longer-term behaviour of the system. To start, we loaded 100 random samples from our posterior into the `rebound` *N*-body integrator (Rein & Liu 2012; Rein & Spiegel 2015). First, we integrated each sample for 100 yr, tracking the resonant arguments $2\lambda_d - \lambda_c - \varpi_c$ and $2\lambda_d - \lambda_c - \varpi_d$. We found that both arguments circulate rather than librate, suggesting that this system is only near and not in resonance despite the small resonant proximity $\Delta = \frac{P_d}{2P_c} - 1 = 5 \times 10^{-3}$. We then integrated 10 samples for 1 Myr, finding them to be stable on these time-scales. Longer integrations were prohibitively expensive as the time-step must be small enough to sample the 3.36 d orbit of the inner planet. We also integrated 1000 random samples for 10 kyr and calculated the distribution for the Mean Exponential Growth of Neighbouring Orbits (MEGNO) parameter (Cincotta & Simó 2000; Rein & Tamayo 2016). We found that the entire sample was tightly clustered around a MEGNO of 2, indicating that the system is not rapidly chaotic. Together, these tests suggest the stability and dynamical regularity of the system.

We then performed a longer time-scale analysis of stability, making use of the Stability Orbital Configuration Klassifier (SPOCK; Tamayo et al. 2020), a machine-learning model capable of classifying the stability of compact 3 + planetary systems with period ratios of adjacent planets $\lesssim 2$ over $10^9$ orbits of the innermost planet, which for our case is $\sim 10^7$ yr. First, we computed the stability of the system by considering the nominal values of the planet parameters derived from our TTV analysis combined with the results of the RV follow-up made by VE21. We used the two models provided by SPOCK: the original machine-learning model FeatureClassier, and DeepRegressor, a Bayesian neural network model (Cranmer et al. 2021). Using these models, we found that the system has a stability probability of 92 per cent and 97 per cent, respectively. Then, once it was established that the system is highly stable in its nominal configuration, we explored the masses and eccentricities of the planets, i.e. the parameters that have the most dramatic effect on the orbital dynamics, by building stability maps of adjacent planets. This was motivated by the fact that perturbation effects among planets fall exponentially with their separation; therefore, non-adjacent interactions can be neglected (Quillen 2011; Petit et al. 2020; Cranmer et al. 2021). The map building entailed exploring the parameter space $M_b$–$M_c$ and $e_b$–$e_c$ for the inner planets b and c, and $M_c$–$M_d$ and $e_c$–$e_d$ for planets c and d. In each case, we explored each parameter within its uncertainty up to 5$\sigma$. In our sets of simulations, we took 10 values from each range, meaning that the size of the stability maps was 10 × 10 pixels. We made use of the two algorithms FeatureClassier and DeepRegressor. In the case of FeatureClassier, for each scenario we set 20 different random initial conditions by varying the longitude of the ascending node and mean anomaly in the range of 0–360 deg. We then averaged these 20 initial conditions to obtain the averaged probability. On the other hand, in the case of DeepRegressor, due to its higher computational cost we set only five different random initial conditions. Hence, for each stability map, we ran 2000 simulations with FeatureClassier and 500 with DeepRegressor. For each stability map, using FeatureClassier and DeepRegressor, respectively, we found the following stability probabilities: (1) for the $M_b$–$M_c$ map, values of $\gtrsim$92 per cent and $\gtrsim$97 per cent; (2) for the $e_b$–$e_c$ map, we found $\gtrsim$70 per cent and $\gtrsim$97 per cent; (3) for the $M_c$–$M_d$ map, we found $\gtrsim$83 per cent and $\gtrsim$97 per cent; and (4) for the $e_c$–$e_d$ map, we found $\gtrsim$86 per cent and $\gtrsim$97 per cent. These results hint that the system is very stable in the 5$\sigma$ range of masses and eccentricities.

## 5 RESULTS AND DISCUSSION

By combining our transit timing results with RV follow-up from VE21, we arrive at precise masses of 1.48 ± 0.18 $M_\oplus$ for planet b, 6.20 ± 0.31 $M_\oplus$ for planet c, and 4.20 ± 0.16 $M_\oplus$ for planet d. We also detect eccentricities for all three planets of $e_b = 0.0167^{+0.0084}_{-0.0089}$, $e_c = 0.0044^{+0.0005}_{-0.0006}$, and $e_d$0.0066 ± 0.0020.

For the star, we find a density of 10.63 ± 0.74 g cm$^{-3}$, in agreement with previous studies. No notable correlation was observed in the stellar parameters being fit, nor was any found in the RV analysis performed by VE21. The final results, including the adopted stellar parameters for all three planets, are summarized in Table 5 We are thus also able to further refine the system's ephemerides, which will be crucial for accurate planning of future transit observations and spectral studies.

### 5.1 Combined TTV and RV analysis

There are only a small handful of exoplanetary systems (about a dozen) for which TTVs and RV data have been analysed both together and separately. In the case of the previously mentioned Kepler 9 system discovered by Holman et al. (2010), which was the first system characterized by TTVs, a series of follow-up studies (Borsato et al. 2014; Borsato et al. 2019) used high-precision RVs in combination with TTVs to better determine the system's masses. While the initial studies reported discrepant masses, with the TTV analysis giving significantly lower masses than those produced by the RV analysis, the 2019 re-analysis with further *Kepler* transits and follow-up with HARPS-North revealed agreement between mass estimates with the two methods (Borsato et al. 2019).

K2-19 is another three-planet, jointly analysed system, which comprised a Neptune-sized inner planet and a Saturn-sized planet in a 2:1 mean-motion resonance, as well as a close-in Earth-sized planet found in K2 data (Armstrong et al. 2015; Nespral et al. 2017). In a similar manner to this study, RV and TTV follow-up of K2-19 by Nespral et al. (2017) demonstrated either approach's ability to constrain planetary masses and further refine them through joint analysis. WASP 47 is a system with four known planets: a hot Jupiter, two sub-Neptunes, and an eccentric, long-period giant planet (Becker et al. 2015). Combining RV data with several years of TTV data allowed for more accurate masses to be obtained for the three inner planets, enabling a clearer picture of these planet's likely two-part evolutionary history to be formed (Weiss et al. 2017).

None the less, several recent studies have noted systematic discrepancies between TTV and RV analyses (Masuda et al. 2013; Weiss et al. 2013). Weiss et al. (2013) noted that for small planets ($<4\,M_\oplus$) TTV studies found systematically lower masses than did RV analyses of the same systems, with Malavolta et al. (2017) positing that such discrepancies could arise from other planets damping the observed effects. Mills & Mazeh (2017) show that these differences are present for planets with periods of greater than 11 d, while shorter-period planets generally find good agreement. Our study points to the compatibility of these two approaches, but more systems with masses analysed jointly by both TTVs and RVs could further illuminate the source of these systematic tensions.

### 5.2 Evolution and additional planets

Systems exhibiting mean-motion resonance can be powerful indicators of evolution. Current models of planetary formation allow for planets to form at any radius, and it is thought that mean-motion resonance develops later on due to dissipative forces, rather than as a direct result of planetary formation. For example, in the two-planet, 2:1 MMR system K2-24, the magnitude of the observed TTVs indicates that disc migration alone could not produce its current configuration, pointing to the role of disc eccentricity damping and excitation, as well as the possible presence of a third planet (Teyssandier & Libert 2020). Mills et al. (2016) demonstrate through the example of the resonant chain in the four-planet Kepler 223 system that migration resulting in often-temporary resonances is likely responsible for the orbits of many close-in sub-Neptune planets.

TTV measurements can also provide significant evidence of further non-transiting planetary companions whose orbits are distributed across inclinations (Brakensiek & Ragozzine 2016). While our results do not cover enough of a super-period to provide direct evidence, they leave open the possibility of an additional planet TOI-270 e either within the orbit of planet b or outside the orbit of planet d. As can be seen in the top panel of Fig. 3, planet b contains several points that are outliers to the main trend line. Individual planet b transits proved difficult to fit and were typically not discernible by eye, so it is likely that these are genuine outliers where the fit was sensitive to airmass and correlated noise. However, it is possible that these points should not be treated as outliers and that further follow-up will reveal a longer super-period trend for this planet, which could be a further indication of a non-transiting companion. With continued monitoring of this system, we can place further constraints on any possible undiscovered companions.

### 5.3 Atmospheric and composition

Our analysis produces densities for the TOI-270 planets of 3.89 ± 0.66 g cm$^{-3}$, 2.70 ± 0.14 g cm$^{-3}$, and 2.90 ± 0.24 g cm$^{-3}$ for

**Table 5.** Stellar and planetary parameters.

| **Parameter** | **Value** | | | **Source** |
|---|---|---|---|---|
| **Star** | **TOI-270**, TIC 259377017, L231-32 | | | |
| Right ascension, declination (J2000) | 04h33m39.72s, −51°57022.44″ | | | |
| Longitude, Latitude (ecl. J2000) | 02h52m35.24s, −71°53049.29″ | | | |
| Magnitudes | $V = 12.62$, TESS = 10.416, $J = 9.099 \pm 0.032$, *Gaia* = 11.63 $H = 8.531 \pm 0.073$, $K = 8.251 \pm 0.029$ | | | |
| Distance, $d_*$ (parsec) | $22.453 \pm 0.021$ | | | |
| Mass, $M_*$ ($M_\odot$) | $0.386 \pm 0.008$ | | | Fit |
| Radius, $R_*$ ($R_\odot$) | $0.380 \pm 0.008$ | | | Fit |
| Density, g cm$^{-3}$ | $10.63 \pm 0.74$ | | | Derived |
| Effective temperature, $T_e$ (K) | $3506 \pm 70$ | | | |
| Surface gravity log ($g$) (cgs) | $4.872 \pm 0.026$ | | | Derived |
| Limb darkening (TESS) $c_1$, $c_2$ | $0.30 \pm 0.04$, $0.22 \pm 0.09$ | | | Fit |
| **Parameter** | **b** | **c** | **d** | **Source** |
| $T_0 - 2\,457\,000$ (BJD) | $1461.01464^{+0.00084}_{-0.00093}$ | $1463.08056 \pm 0.00040$ | $1469.33823 \pm 0.00032$ | Fit |
| Period (days) | $3.35992 \pm 0.00005$ | $5.66051 \pm 0.00004$ | $11.38194 \pm 0.00011$ | Fit |
| $R_p/R_*$ | $0.0307^{+0.0012}_{-0.0009}$ | $0.0560 \pm 0.0017$ | $0.0491 \pm 0.0016$ | Derived |
| cos $i$ | $0.0267^{+0.0256}_{-0.0184}$ | $0.01393^{+0.00230}_{-0.00282}$ | $0.0062^{+0.0019}_{-0.0026}$ | Fit |
| $a/R_*$ | $17.108^{+3.526}_{-1.858}$ | $25.569 \pm 0.306$ | $41.744 \pm 0.527$ | Derived |
| Mass $M_\oplus$ | $1.48 \pm 0.18$ | $6.20 \pm 0.31$ | $4.20 \pm 0.16$ | Fit |
| Radius $R_\oplus$ | $1.28^{+0.05}_{-0.04}$ | $2.33 \pm .01$ | $2.00 \pm .05$ | Fit |
| Eccentricity $e$ | $0.0167^{+0.0084}_{-0.0089}$ | $0.0044^{+0.0005}_{-0.0006}$ | $0.0066 \pm 0.0020$ | Fit |
| Impact parameter $b$ | $0.48 \pm 0.46$ | $0.35 \pm 0.06$ | $0.23 \pm 0.08$ | Derived |
| Transit dur (hours) | $1.294 \pm 0.026$ | $1.682 \pm 0.019$ | $2.117 \pm 0.018$ | Derived |
| Density $\rho_p$ (g cm$^{-3}$) | $3.89 \pm 0.66$ | $2.70 \pm 0.14$ | $2.90 \pm 0.24$ | Derived |
| $T_{eq}$ (albedo = 0.3) | $548 \pm 15$ | $448 \pm 10$ | $351 \pm 8$ | Derived |
| $T_{eq}$ (albedo = 0) | $600 \pm 16$ | $489 \pm 11$ | $383 \pm 9$ | Derived |
| Surface gravity $\log_{10}(g)$ (cgs) | $2.95 \pm 0.43$ | $3.05 \pm 0.15$ | $3.01 \pm 0.19$ | Derived |

*Note.* Stellar parameters are drawn from VE21 and G20. Both fit and derived planetary parameters from this study are shown. Timing fits were performed by fixing planetary parameters to the *Spitzer*-derived values for planet c, and for *TESS* all-sector values for planets b and d.

planets b, c, and, d, respectively. Their mass–radius relation places them in the region of super-Earth and sub-Neptunes, as seen in Fig. 4. The innermost planet has a significantly higher density making its composition likely rocky and Earth-like. To date, TOI-270 b is one of the smallest planets to be characterized with joint TTV and RV analysis. TOI-270 c and TOI-270 d have lower densities, which suggests similarly rocky core composition with the addition of a more substantial H/He atmosphere. Compared to VE21, our adopted radii change the densities for planet c and d slightly. As seen in Fig. 4, this does not alter the bulk planetary characteristics significantly but shifts the atmospheric composition, increasing the ratio of $H_2$ to $H_2O$.

### 5.4 Potential for atmospheric characterization

Planets between the size of Earth and Neptune are very common results of planet formation, and it appears that approximately half of Sun-like stars have at least one sub-Neptune in an orbit closer than that of Mercury (Petigura, Howard & Marcy 2013). However, no such planets exist in our Solar system and our understanding of their evolution and atmospheres remains limited. One of the best studied sub-Neptunes to date is GJ 1214 b, which has a very flat transmission spectrum, most likely because of a high-altitude haze formed through methane photochemistry (Kreidberg et al. 2014).

Cool sub-Neptunes, in particular, are useful probes of planet formation. Because they have not been exposed to such extreme photo-evaporation as their more heavily irradiated counterparts, they retain more of their initial atmospheres. In the case where the planet is relatively free from clouds or haze, this can be paired with spectroscopic observations to provide excellent evidence for different models of planet formation. Even with haze present, as is the case for GJ 1214 b, and simulated analogues, Hood et al. (2021) demonstrate that with high-resolution infrared transmission spectra, numerous molecular features (such as for $H_2O$, CO, and $CO_2$) can be detected in hazy sub-Neptune spectra that otherwise appear ‘featureless’, with even more promise at higher resolution with upcoming ELT instruments. White, Mishra & Lewis (2021) further demonstrate the potential of JWST to penetrate through hazy layers and detect features of $CO_2$ and $CH_4$ for GJ 1214 b,.

We calculated the transmission spectra metric (TSM) for the planets of TOI-270 as described by Kempton et al. (2018), which is meant to identify top *TESS* targets for atmospheric characterization defining an expected signal to noise for a given planet’s transmission and emission spectroscopy, and which enables us to compare the amenability to atmospheric follow-up of our system with that of others. All three targets fall well above the suggested cutoff for follow-up and among the highest (most promising) for small-sized targets below 600K. Planet b, which is classed by Kempton et al. (2018) as terrestrial, has a TSM of 102.1, above the suggested TSM cutoff of 12. Planets c and d, which are classed as small sub-Neptunes, have TSMs of 125.8 and 124.0, respectively, above the suggested cutoff of 92.

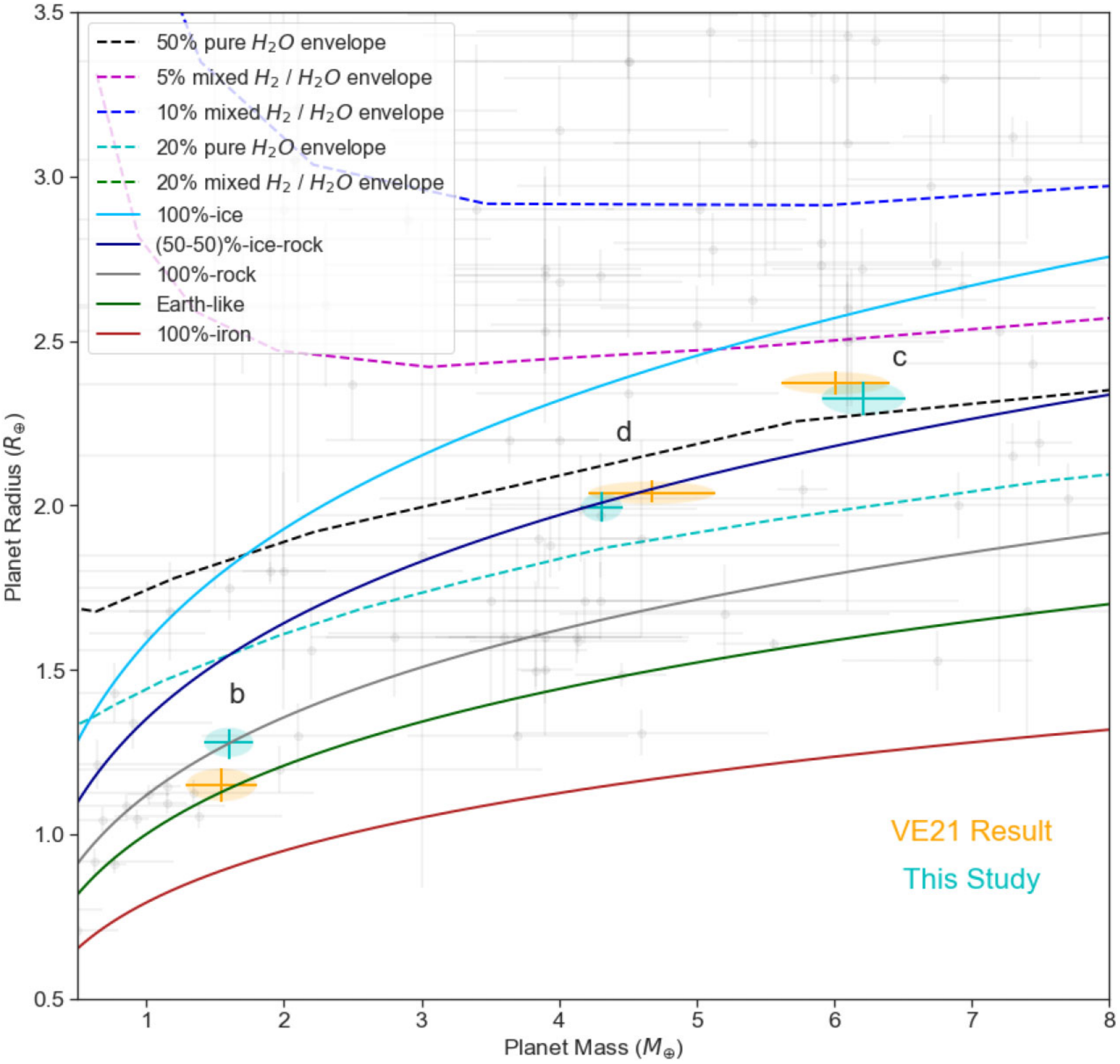

**Figure 4.** Mass fits shown against a population of small mass and radius known exoplanets obtained from the NASA Exoplanet Archive (in grey). The orange points represents the RV mass priors from VE21, which are consistent with the posterior to two-sigma level, and the teal points show the results of our TTV fit. Because our TTV fitting does not incorporate planetary radius, these two parameters are not correlated in our study. The dotted lines describe gas envelopes of varying composition. All composition profiles are from Kopparapu et al. (2014) (water/ice: $H_2O$; rock: $Mg_2SiO4$; iron: Fe; Earth-like: 67 per cent rock/33 per cent iron).

## 6 CONCLUSION AND FUTURE WORK

We report 2 yr of follow-up transit observations of the three TOI-270 planets, detecting significant TTVs, which enable us to refine planetary parameters, masses, and eccentricities.

This work demonstrates the potential of TTV and joint TTV-RV analyses to refine the orbital parameters and masses of small near-resonant planets to within 5–10 per cent. Systems that have been jointly analysed by these approaches are limited, and as *TESS* re-observes targets from previous cycles in the coming years, further such studies could serve to emphasize the strengths of this combination as well as help elucidate potential discrepancies and systematic biases.

Our work took a self-consistent approach to light-curve analysis between eight different observatories, including reanalysing *TESS* data with the inclusion of new observations from cycle 3, finding largely consistent refinements to planetary parameters and orbital ephemerides found by G20 and VE21, and producing a picture of a dynamically stable system of super-Earth/sub-Neptune planets, which show significant promise for future comparative planetology through atmospheric studies. TOI-270 will continue to be observed through 2021, with transit studies done with *LCO* (program id *LCO* 2021A, PI Parviainen), *ASTEP*, *NGTS*, and *TRAPPIST*, which will enable further TTV characterization.

Ongoing observations of TOI-270 with *HST* (program id GO-15814, PI Mikal-Evans) should provide a valuable glimpse into the system's chemical abundances, as well as a unique opportunity for comparison between the two outer planets, whose similar size and density lend itself extremely well to comparative planetology. Of particular interest is the methane content of each planet, which is expected to be measured to high sensitivity, as is the presence or absence of photochemical haze on either or both of the planets, which would allow for a direct comparison of the effect of temperature and irradiation on haze production.

Obtaining accurate masses is critical to the interpretation of follow-up atmospheric studies and evolutionary models and provides further insight into the benefits of joint TTV and RV analyses.

## ACKNOWLEDGEMENTS

This research received funding from the European Research Council (ERC) under the European Union's Horizon 2020 research and innovation programme (grant agreement no. 803193/BEBOP) and from the Science and Technology Facilities Council (STFC; grant no. ST/S00193X/1). LK acknowledges funding support from the Clarendon Scholarship. SV is supported by an NSF Graduate Research Fellowship and the Paul & Daisy Soros Fellowship for New Americans, and additionally thanks Heather Knutson for helpful comments on the dynamical analysis. MNG acknowledges support from MIT's Kavli Institute as a Juan Carlos Torres Fellow. TD acknowledges support from MIT's Kavli Institute as a Kavli Fellow. DD acknowledges support from the tess Guest Investigator Program grant 80NSSC19K1727 and NASA Exoplanet Research Program grant 18-2XRP182-0136. JSJ acknowledges support by FONDECYT grant 1201371, and partial support from CONICYT project Basal AFB-170002. JIV acknowledges support of CONICYT-PFCHA/Doctorado Nacional-21191829. ACh acknowledges the support of the DFG priority program SPP 1992 'Exploring the Diversity of Extrasolar Planets' (RA 714/13-1), DJA acknowledges support from the STFC via an Ernest Rutherford Fellowship (ST/R00384X/1) JSJ acknowledges support by FONDECYT grant 1201371, and partial support from CONICYT project Basal AFB-170002. PJW is supported by STFC consolidated grant ST/T000406/1. SG has been supported by STFC through consolidated grants ST/L000733/1 and ST/P000495/1. SLC acknowledges the support of an STFC Ernest Rutherford Fellowship ST/R003726/1. JIV acknowledges support of CONICYT-PFCHA/Doctorado Nacional-21191829.

This paper contributes to meeting the TESS Mission Level One Science Requirement: 'The TESS team shall assure that the masses of fifty (50) planets with radii less than 4 REarth are determined.' Funding for the TESS mission is provided by NASA's Science Mission Directorate. We acknowledge the use of public TESS data from pipelines at the TESS Science Office and at the TESS Science Processing Operations Center. This research has made use of the Exoplanet Follow-up Observation Program website, which is operated by the California Institute of Technology, under contract with the National Aeronautics and Space Administration under the Exoplanet Exploration Program. Resources supporting this work were provided by the NASA High-End Computing (HEC) Program through the NASA Advanced Supercomputing (NAS) Division at Ames Research Center for the production of the SPOC data products. This paper includes data collected by the TESS mission that are publicly available from the Mikulski Archive for Space Telescopes (MAST).

Resources supporting this work were provided by the NASA High-End Computing (HEC) Program through the NASA Advanced Supercomputing (NAS) Division at Ames Research Center for the production of the SPOC data products. This work makes use of observations from the LCOGT network. Part of the LCOGT telescope time was granted by NOIRLab through the Mid-Scale Innovations Program (MSIP). MSIP is funded by NSF. The research leading to these results has received funding from the ARC grant for Concerted Research Actions, financed by the Wallonia-Brussels Federation. TRAPPIST is funded by the Belgian Fund for Scientific Research (Fond National de la Recherche Scientifique, FNRS) under the grant FRFC 2.5.594.09.F, with the participation of the Swiss National Science Foundation (SNF). MG and EJ are F.R.S.-FNRS Senior Research Associate. ASTEP benefited from the support of the French and Italian polar agencies IPEV and PNRA in the framework of the Concordia station program. TG, AA, LA, DM, and F-XS acknowledge support from Idex UCAJEDI (ANR-15-IDEX-01). Based on data collected under the NGTS project at the ESO La Silla Paranal Observatory. The NGTS facility is operated by the consortium institutes with support from the UK Science and Technology Facilities Council (STFC) projects ST/M001962/1 and ST/S002642/1. This work has made use of data from the European Space Agency (ESA) mission *Gaia* (https://www.cosmos.esa.int/gaia), processed by the *Gaia* Data Processing and Analysis Consortium (DPAC, https://www.cosmos.esa.int/web/gaia/dpac/consortium). Funding for the DPAC has been provided by national institutions, in particular, the institutions participating in the *Gaia* Multilateral Agreement. This work has been carried out in the framework of the PlanetS National Centres of Competence in Research (NCCR) supported by the Swiss National Science Foundation (SNSF).

## DATA AVAILABILITY

The *TESS* data used within this study are hosted and made publicly available by the Mikulski Archive for Space Telescopes (MAST, http://archive.stsci.edu/tess/). LCO, NGTS, and PEST observations are through Exoplanet Follow-up Observing Program for TESS (ExoFOP-TESS) website as community TOIs (cTOIs; https://exofop.ipac.caltech.edu/tess/target.php?id = 259377017).

The models and analyses of transit timings were conducted using publicly available open software codes, ASTROPY (Astropy Collaboration 2013), AstroImageJ (Collins et al. 2017), BATMAN (Kreidberg 2015), EMCEE (Foreman-Mackey et al. 2013, 2019), GEORGE (Ambikasaran et al. 2014), IDL (Landsman 1993), LDTK (Parviainen & Aigrain 2015), Tapir (Jensen 2013), and TTVFast (Deck et al. 2014)

Facilities used: CTIO:1.3 m, CTIO:1.5 m, CXO, Las Cumbres Observatory Global Telescope (LCOGT), TESS, *Spitzer Space Telescope*, ASTEP, NGTS, and TRAPPIST.

## APPENDIX A:

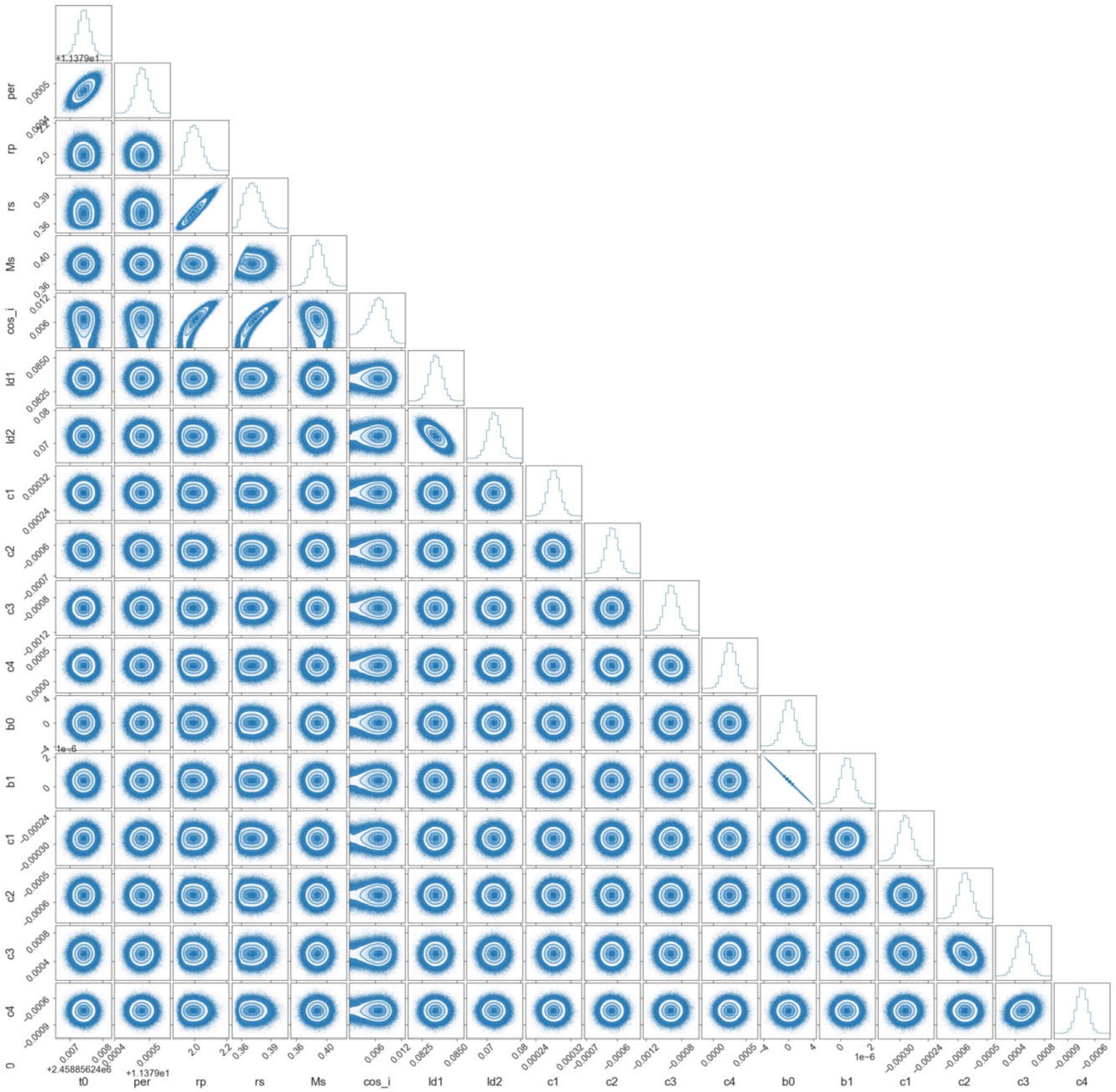

**Figure A1.** Corner plot showing the posterior distribution for a representative subset of parameters of the simultaneous Spitzer transit fit for planet d, incorporating three transits. The fit uses pixel-level decorrelation (PLD), as described by Section 3.3.

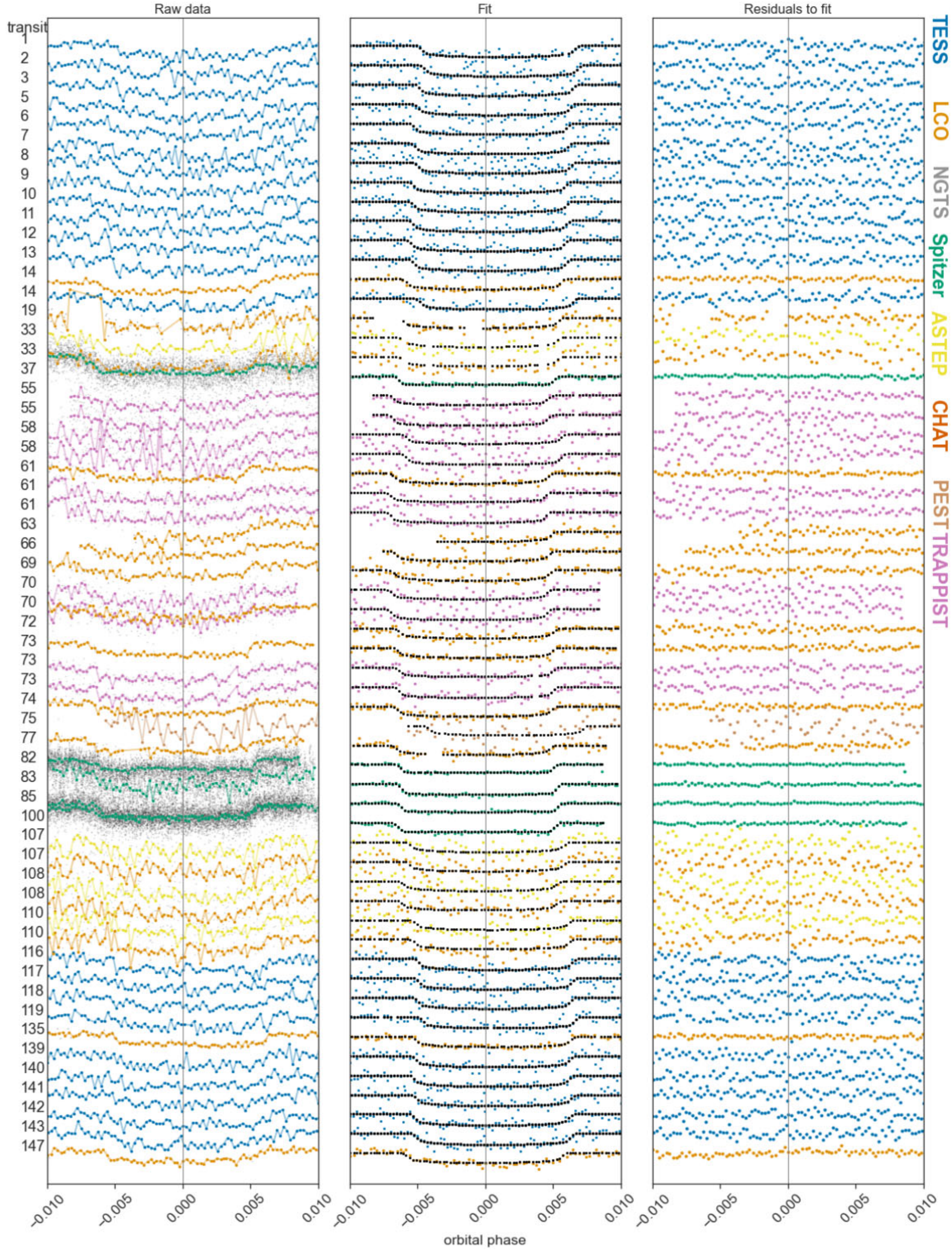

**Figure A2.** Planet c transits. The leftmost column shows unprocessed light curves with data binned to *TESS* 2-min cadence for clarity. The middle column shows the fit by the GP plus the transit curve and the rightmost column gives the residuals to this fit. As can be seen in some cases, there were not enough data points to constrain the fitting. These observations were subsequently discarded before analysis.

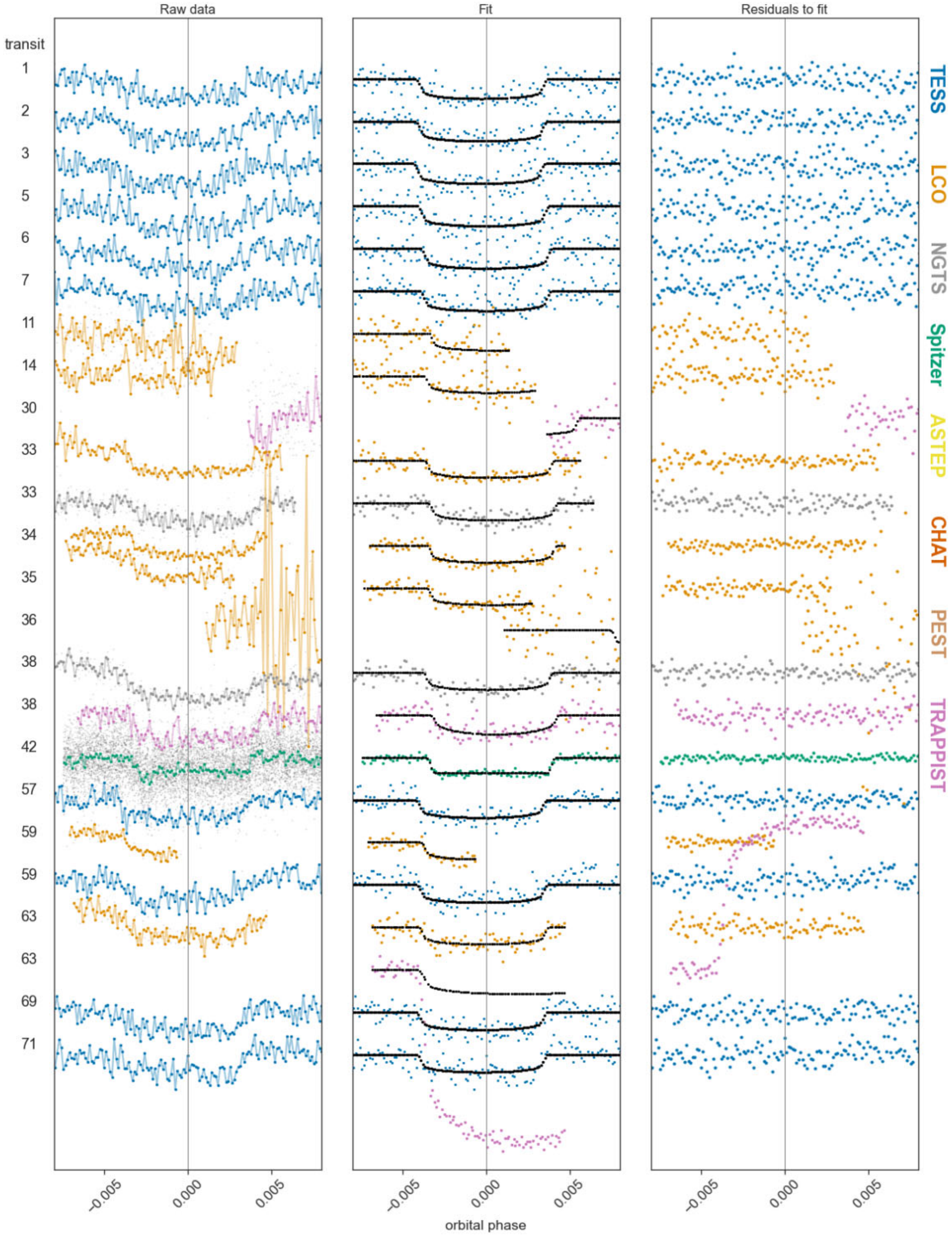

**Figure A3.** Planet d transits. Similar to Fig. A2, each colour represents the respective telescope. Of note, several transits resulted in visibly incorrect fits, due to incomplete transit coverage or remnant correlated noise (most often due to high airmass). These transits were excluded from further analysis.

**Table A1.** Transit timing variations planet b.

| Transit number | Telescope | Mid-transit time (BJD | Timing offset (days) | Error |
|---|---|---|---|---|
| **TOI-270 b** | | | | |
| 1 | TESS | 2458387.09027 | −0.00261 | $^{+0.02110}_{-0.00899}$ |
| 2 | TESS | 2458390.45207 | −0.00089 | $^{+0.00378}_{-0.00719}$ |
| 3 | TESS | 2458393.81351 | 0.00047 | $^{+0.01845}_{-0.01340}$ |
| 4 | TESS | 2458397.16906 | −0.00406 | $^{+0.02853}_{-0.01907}$ |
| 5 | TESS | 2458400.53171 | −0.00149 | $^{+0.01164}_{-0.00563}$ |
| 6 | TESS | 2458403.89026 | −0.00302 | $^{+0.00561}_{-0.00158}$ |
| 9 | TESS | 2458413.98191 | 0.00839 | $^{+0.00724}_{-0.00195}$ |
| 10 | TESS | 2458417.33404 | 0.00044 | $^{+0.00937}_{-0.00525}$ |
| 13 | TESS | 2458427.42528 | 0.01144 | $^{+0.00135}_{-0.01211}$ |
| 15 | TESS | 2458434.13850 | 0.00450 | $^{+0.00714}_{-0.00048}$ |
| 17 | TESS | 2458440.85173 | −0.00243 | $^{+0.02183}_{-0.00992}$ |
| 18 | TESS | 2458444.21671 | 0.00247 | $^{+0.00253}_{-0.00687}$ |
| 19 | TESS | 2458447.57362 | −0.00070 | $^{+0.00130}_{-0.00304}$ |
| 22 | TESS | 2458457.65648 | 0.00192 | $^{+0.00920}_{-0.00671}$ |
| 23 | TESS | 2458461.01628 | 0.00164 | $^{+0.01262}_{-0.00072}$ |
| 34 | LCO | 2458497.97113 | −0.00439 | $^{+0.01585}_{-0.00122}$ |
| 37 | LCO | 2458508.05202 | −0.00374 | $^{+0.01428}_{-0.01056}$ |
| 38 | LCO | 2458511.41210 | −0.00374 | $^{+0.00679}_{-0.00338}$ |
| 103 | NGTS | 2458729.82844 | 0.00740 | $^{+0.00311}_{-0.01621}$ |
| 133 | LCO | 2458830.63207 | 0.00863 | $^{+0.01162}_{-0.01385}$ |
| 215 | LCO | 2459106.16860 | 0.01860 | $^{+0.00022}_{-0.00555}$ |
| 218 | TESS | 2459116.23825 | 0.00801 | $^{+0.01181}_{-0.00759}$ |
| 219 | TESS | 2459119.60599 | 0.01567 | $^{+0.00526}_{-0.00555}$ |
| 220 | TESS | 2459122.95868 | 0.00828 | $^{+0.01197}_{-0.01316}$ |
| 221 | TESS | 2459126.32468 | 0.01420 | $^{+0.00414}_{-0.00683}$ |
| 223 | TESS | 2459133.04645 | 0.01581 | $^{+0.01880}_{-0.02103}$ |
| 224 | TESS | 2459136.40990 | 0.01918 | $^{+0.01333}_{-0.03099}$ |
| 225 | TESS | 2459139.76388 | 0.01308 | $^{+0.01296}_{-0.02113}$ |
| 227 | LCO | 2459146.48644 | 0.01548 | $^{+0.00794}_{-0.01916}$ |
| 236 | TESS | 2459176.73154 | 0.01986 | $^{+0.00112}_{-0.00293}$ |
| 237 | TESS | 2459180.08965 | 0.01789 | $^{+0.00268}_{-0.00114}$ |
| 238 | TESS | 2459183.45059 | 0.01875 | $^{+0.00004}_{-0.00585}$ |
| 240 | TESS | 2459190.16616 | 0.01416 | $^{+0.00736}_{-0.00111}$ |
| 241 | TESS | 2459193.53089 | 0.01881 | $^{+0.01788}_{-0.02524}$ |
| 242 | TESS | 2459196.88738 | 0.01522 | $^{+0.00599}_{-0.00208}$ |

*Note.* Mid-Transit times for planet b. The Timing Offset is reported in relation to the initial G20 *TESS* linear ephemerides.

**Table A2.** Transit timing variations planet c.

| Transit number | Telescope | Mid-transit time (BJD | Timing offset (days) | Error |
|---|---|---|---|---|
| **TOI-270 c** | | | | |
| 0 | TESS | 2458383.84177 | 0.00376 | $^{+0.00086}_{-0.00093}$ |
| 1 | TESS | 2458389.50354 | 0.00503 | $^{+0.00098}_{-0.00141}$ |
| 2 | TESS | 2458395.16239 | 0.00340 | $^{+0.00064}_{-0.00060}$ |
| 4 | TESS | 2458406.48248 | 0.00249 | $^{+0.00106}_{-0.00134}$ |
| 5 | TESS | 2458412.14178 | 0.00131 | $^{+0.00073}_{-0.00091}$ |
| 6 | TESS | 2458417.80261 | 0.00164 | $^{+0.00145}_{-0.00126}$ |
| 7 | TESS | 2458423.46444 | 0.00298 | $^{+0.00084}_{-0.00081}$ |
| 8 | TESS | 2458429.12270 | 0.00075 | $^{+0.00103}_{-0.00098}$ |
| 9 | TESS | 2458434.78365 | 0.00120 | $^{+0.00101}_{-0.00097}$ |
| 10 | TESS | 2458440.44412 | 0.00118 | $^{+0.00127}_{-0.00155}$ |
| 11 | TESS | 2458446.10403 | 0.00060 | $^{+0.00087}_{-0.00094}$ |
| 12 | TESS | 2458451.76531 | 0.00138 | $^{+0.00128}_{-0.00114}$ |
| 13 | TESS | 2458457.42794 | 0.00352 | $^{+0.00394}_{-0.00205}$ |
| 14 | LCO | 2458463.08389 | −0.00103 | $^{+0.00032}_{-0.00033}$ |
| 19 | LCO | 2458491.38544 | −0.00193 | $^{+0.00037}_{-0.00043}$ |
| 33 | ASTEP | 2458570.63326 | −0.00102 | $^{+0.00149}_{-0.00134}$ |
| 37 | Spitzer | 2458593.27348 | −0.00277 | $^{+0.00018}_{-0.00018}$ |
| 55 | TRAPPIST | 2458695.16181 | −0.00332 | $^{+0.00061}_{-0.00085}$ |
| 61 | TRAPPIST | 2458729.12413 | −0.00395 | $^{+0.00067}_{-0.00066}$ |
| 61 | LCO | 2458729.12540 | −0.00268 | $^{+0.00036}_{-0.00036}$ |
| 63 | LCO | 2458740.44769 | −0.00138 | $^{+0.00056}_{-0.00053}$ |
| 66 | LCO | 2458757.42889 | −0.00166 | $^{+0.00031}_{-0.00031}$ |
| 69 | LCO | 2458774.41028 | −0.00175 | $^{+0.00040}_{-0.00038}$ |
| 70 | TRAPPIST | 2458780.06961 | −0.00291 | $^{+0.00080}_{-0.00096}$ |
| 72 | LCO | 2458791.39388 | 0.00037 | $^{+0.00033}_{-0.00032}$ |
| 73 | LCO | 2458797.05387 | −0.00013 | $^{+0.00034}_{-0.00035}$ |
| 73 | TRAPPIST | 2458797.05430 | 0.00030 | $^{+0.00066}_{-0.00067}$ |
| 74 | LCO | 2458802.71417 | −0.00032 | $^{+0.00034}_{-0.00036}$ |
| 77 | LCO | 2458819.69528 | −0.00069 | $^{+0.00128}_{-0.00100}$ |
| 78 | Spitzer | 2458825.35745 | 0.00098 | $^{+0.00019}_{-0.00018}$ |
| 81 | Spitzer | 2458842.33935 | 0.00141 | $^{+0.00014}_{-0.00014}$ |
| 83 | Spitzer | 2458853.66120 | 0.00228 | $^{+0.00015}_{-0.00015}$ |
| 84 | Spitzer | 2458859.32203 | 0.00261 | $^{+0.00013}_{-0.00012}$ |
| 107 | ASTEP | 2458989.51812 | 0.00736 | $^{+0.00106}_{-0.00113}$ |
| 108 | ASTEP | 2458995.17581 | 0.00455 | $^{+0.00200}_{-0.00195}$ |
| 110 | ASTEP | 2459006.50192 | 0.00968 | $^{+0.00105}_{-0.00106}$ |
| 129 | TESS | 2459114.05383 | 0.01222 | $^{+0.00070}_{-0.00072}$ |
| 130 | TESS | 2459119.71415 | 0.01205 | $^{+0.00099}_{-0.00100}$ |
| 131 | TESS | 2459125.37407 | 0.01147 | $^{+0.00076}_{-0.00076}$ |
| 132 | TESS | 2459131.03666 | 0.01358 | $^{+0.00069}_{-0.00071}$ |
| 135 | LCO | 2459148.01698 | 0.01241 | $^{+0.00027}_{-0.00029}$ |
| 139 | TESS | 2459170.65889 | 0.01235 | $^{+0.00086}_{-0.00090}$ |
| 140 | TESS | 2459176.31907 | 0.01204 | $^{+0.00081}_{-0.00079}$ |
| 141 | TESS | 2459181.98058 | 0.01305 | $^{+0.00100}_{-0.00101}$ |
| 142 | TESS | 2459187.64029 | 0.01228 | $^{+0.00092}_{-0.00088}$ |
| 143 | TESS | 2459193.30171 | 0.01320 | $^{+0.00129}_{-0.00124}$ |
| 147 | LCO | 2459215.94006 | 0.00958 | $^{+0.00112}_{-0.00111}$ |

*Note.* Mid-transit times for planet c. The Timing offset is reported in relation to the initial G20 *TESS* linear ephemerides.

**Table A3.** Transit timing variations planet d.

| Transit number | Telescope | Mid-transit time (BJD | Timing offset (days) | Error |
|---|---|---|---|---|
| **TOI-270 d** | | | | |
| 1 | TESS | 2458389.67893 | −0.00320 | $^{+0.00118}_{-0.00097}$ |
| 2 | TESS | 2458401.05791 | −0.00370 | $^{+0.00118}_{-0.00158}$ |
| 3 | TESS | 2458412.43573 | −0.00536 | $^{+0.00614}_{-0.00128}$ |
| 5 | TESS | 2458435.19786 | −0.00219 | $^{+0.00087}_{-0.00098}$ |
| 6 | TESS | 2458446.57953 | 0.00000 | $^{+0.00203}_{-0.00164}$ |
| 7 | TESS | 2458457.95868 | −0.00033 | $^{+0.00449}_{-0.00243}$ |
| 14 | LCO | 2458537.61795 | 0.00259 | $^{+0.00067}_{-0.00075}$ |
| 24 | Spitzer | 2458651.41688 | 0.00672 | $^{+0.00025}_{-0.00024}$ |
| 33 | LCO | 2458753.83030 | 0.00484 | $^{+0.00033}_{-0.00033}$ |
| 33 | NGTS | 2458753.83266 | 0.00720 | $^{+0.00108}_{-0.00125}$ |
| 34 | LCO | 2458765.21150 | 0.00656 | $^{+0.00046}_{-0.00047}$ |
| 35 | LCO | 2458776.59113 | 0.00671 | $^{+0.00071}_{-0.00069}$ |
| 38 | NGTS | 2458810.72707 | 0.00421 | $^{+0.00079}_{-0.00083}$ |
| 38 | TRAPPIST | 2458810.73193 | 0.00908 | $^{+0.00142}_{-0.00116}$ |
| 39 | Spitzer | 2458822.10886 | 0.00652 | $^{+0.00023}_{-0.00022}$ |
| 42 | Spitzer | 2458856.24751 | 0.00674 | $^{+0.00024}_{-0.00023}$ |
| 59 | LCO | 2459049.69684 | 0.00492 | $^{+0.00070}_{-0.00070}$ |
| 63 | LCO | 2459095.21369 | 0.00386 | $^{+0.00043}_{-0.00037}$ |
| 65 | TESS | 2459117.97101 | 0.00221 | $^{+0.00094}_{-0.00107}$ |
| 67 | TESS | 2459140.73102 | 0.00327 | $^{+0.00118}_{-0.00128}$ |
| 70 | TESS | 2459174.86949 | 0.00331 | $^{+0.00136}_{-0.00119}$ |
| 72 | TESS | 2459197.62899 | 0.00385 | $^{+0.00123}_{-0.00106}$ |

*Note.* Mid-transit times for planet d. The timing offset is reported in relation to the initial G20 *TESS* linear ephemerides.

[1] *Department of Physics, University of Oxford, Keble Road, Oxford OX1 3RH, UK*
[2] *Division of Geological and Planetary Sciences, California Institute of Technology, 1200 East California Blvd, Pasadena, CA 91125, USA*
[3] *Department of Physics, Massachusetts Institute of Technology, Cambridge, MA 02139, USA*
[4] *Juan Carlos Torres Fellow*
[5] *Kavli Institute for Astrophysics and Space Research, Massachusetts Institute of Technology, Cambridge, MA 02139, USA*
[6] *Department of Physics & Astronomy, Swarthmore College, Swarthmore, PA 19081, USA*
[7] *Instituto de Astrofísica de Canarias (IAC), E-38200 La Laguna, Tenerife, Spain*
[8] *Department Astrofísica, Universidad de La Laguna (ULL), E-38206 La Laguna, Tenerife, Spain*
[9] *Space Sciences, Technologies and Astrophysics Research (STAR) Institute, Université de Liège, 19C Allée du 6 Août, B-4000 Liège, Belgium*
[10] *Astrobiology Research Unit, Université de Liège, 19C Allée du 6 Août, B-4000 Liège, Belgium*
[11] *Université Côte d'Azur, Observatoire de la Côte d'Azur, CNRS, Laboratoire Lagrange, Bd de l'Observatoire, CS 34229, F-06304 Nice cedex 4, France*
[12] *School of Physics and Astronomy, University of Leicester, University Road, Leicester LE1 7RH, UK*
[13] *Departamento de Astronomía, Universidad de Chile, Casilla 36-D, Santiago, Chile*
[14] *Centre for Exoplanets and Habitability, University of Warwick, Gibbet Hill Road, Coventry CV4 7AL, UK*
[15] *Department of Physics, University of Warwick, Gibbet Hill Road, Coventry CV4 7AL, UK*
[16] *Oukaimeden Observatory, High Energy Physics and Astrophysics Laboratory, Cadi Ayyad University, Marrakech, Morocco*
[17] *Department of Physics and Institute for Research on Exoplanets, Université de Montréal, Montreal, QC, Canada*
[18] *Astrophysics Science Division, NASA Goddard Space Flight Center, Greenbelt, MD 20771, USA*
[19] *Facultad de Ingeniería y Ciencias, Universidad Adolfo Ibáñez, Av. Diagonal las Torres 2640, Peñalolén, Santiago, Chile*
[20] *Millennium Institute for Astrophysics, 4860, Macul, Santiago, Chile*
[21] *Concordia Station, IPEV/PNRA, Antarctica*
[22] *Department of Physics and Astronomy, University of Leicester, Leicester LE1 7RH, UK*
[23] *Caltech/IPAC-NASA Exoplanet Science Institute Pasadena, CA 91125, USA*
[24] *Banting Fellow*
[25] *Center for Astrophysics | Harvard & Smithsonian, 60 Garden Street, Cambridge, MA 02138, USA*
[26] *George Mason University, 4400 University Drive, Fairfax, VA 22030 USA*
[27] *American Association of Variable Star Observers, 49 Bay State Road, Cambridge, MA 02138, USA*
[28] *Department of Physics and Astronomy, University of Kansas, Pasadena, CA 91125, USA*
[29] *European Space Agency (ESA), European Space Research and Technology Centre (ESTEC), Keplerlaan 1, NL-2201 AZ Noordwijk, the Netherlands*
[30] *Kavli Fellow*
[31] *Department of Physics and Astronomy, University of New Mexico, 210 Yale Blvd NE, Albuquerque, NM 87106, USA*
[32] *School of Physics & Astronomy, University of Birmingham, Edgbaston, Birmingham B15 2TT, UK*
[33] *Department of Astronomy and Astrophysics, University of Chicago, 5640 S Ellis Ave Chicago, IL 60637, USA*
[34] *Department of Astronomy and Tsinghua Centre for Astrophysics, Tsinghua University, Beijing 100084, China*

[35]*Jet Propulsion Laboratory, California Institute of Technology, 4800 Oak Grove Drive, Pasadena, CA 91109, USA*
[36]*Núcleo de Astronomía, Facultad de Ingeniería y Ciencias, Universidad Diego Portales, Av. Ejército 441, Santiago, Chile*
[37]*Observatoire Astronomique de l'Université de Genève, Chemin Pegasi CH-51, Versoix, Switzerland*
[38]*John F. Kennedy High School, 3000 Bellmore Avenue, Bellmore, NY 11710, USA*
[39]*Department of Earth and Planetary Sciences, University of California, Riverside, CA 92521, USA*
[40]*Department of Physics and Astronomy, University of Louisville, Louisville, KY 40292, USA*
[41]*Instituto de Astronomía, Universidad Católica del Norte, 1270709, Antofagasta, Chile*
[42]*Patashnick Voorheesville Observatory, Voorheesville, NY 12186, USA*
[43]*Department of Earth, Atmospheric and Planetary Sciences, Massachusetts Institute of Technology, Cambridge, MA 02139, USA*
[44]*Department of Aeronautics and Astronautics, MIT, 77 Massachusetts Avenue, Cambridge, MA 02139, USA*
[45]*Institute of Planetary Research, German Aerospace Center, Rutherfordstraße 2, D-12489 Berlin, Germany*
[46]*Perth Exoplanet Survey Telescope, Perth, Western, WA 6076, Australia*
[47]*SETI Institute, Mountain View, CA 94043, USA*
[48]*NASA Ames Research Center, Moffett Field, CA 94035, USA*
[49]*Tsinghua International School, Beijing 100084, China*
[50]*Stanford Online High School, 415 Broadway Academy Hall, Floor 2, 8853, Redwood City, CA 94063, USA*
[51]*Department of Astrophysical Sciences, Princeton University, 4 Ivy Lane, Princeton, NJ 08544, USA*

*This paper has been typeset from a TEX/LATEX file prepared by the author.?>*